\documentclass[sn-basic]{sn-jnl}

\usepackage{graphicx}%
\usepackage{aas_macros}
\usepackage{multirow}%
\usepackage{amsmath,amssymb,amsfonts}%
\usepackage{amsthm}%
\usepackage{mathrsfs}%
\usepackage[title]{appendix}%
\usepackage{xcolor}%
\usepackage{textcomp}%
\usepackage{manyfoot}%
\usepackage{booktabs}%
\usepackage{algorithm}%
\usepackage{algorithmicx}%
\usepackage{algpseudocode}%
\usepackage{listings}%
\usepackage{lineno}

\theoremstyle{thmstyleone}%
\theoremstyle{thmstyletwo}%

\theoremstyle{thmstylethree}%

\begin{document}

\title[Article Title]{HyperScout-H: the hyperspectral imager for the ESA Hera mission}


\author*[1,2]{\fnm{Marcel M.} \sur{Popescu}}\email{popescu.marcel@ucv.ro}
\equalcont{These authors contributed equally to this work.}
\author[3,4]{\fnm{Julia} \sur{de Le\'{o}n}}
\equalcont{These authors contributed equally to this work.}
\author[1,3,4]{\fnm{George Pantelimon} \sur{Prodan}}
\author[5]{\fnm{Michael} \sur{K\"{u}ppers}}
\author[6]{\fnm{G\'{a}bor} \sur{Kov\'{a}cs}}
\author[6]{\fnm{Bal\'{a}zs Vince} \sur{Nagy}} 
\author[7]{\fnm{Bj\"{o}rn} \sur{Grieger}}
\author[8]{\fnm{Hannah} \sur{Goldberg}}
\author[9]{\fnm{Marco} \sur{Esposito}}
\author[9]{\fnm{Nathan} \sur{Vercruyssen}}
\author[10]{\fnm{Eri} \sur{Tatsumi}}
\author[11]{\fnm{Lisa} \sur{Kr\"{a}mer Ruggiu}}
\author[12]{\fnm{\"{O}zg\"{u}r} \sur{Karatekin}}
\author[13]{\fnm{Seiji} \sur{Sugita}}
\author[14]{\fnm{Monica} \sur{Lazzarin}}
\author[15]{\fnm{Paul A.} \sur{Abell}}
\author[16]{\fnm{Jean-Baptiste} \sur{Vincent}}
\author[1]{\fnm{Iulian} \sur{Petri\c{s}or}}
\author[17]{\fnm{Patrick} \sur{Michel}}

\affil*[1]{\orgname{University of Craiova}, \orgaddress{\street{Strada Alexandru Ioan Cuza 13}, \city{Craiova}, \postcode{200585}, \country{Romania}}}

\affil[2]{\orgname{Institute of Space Science (ISS)}, \orgaddress{\street{409 Atomi\c{s}tilor Street}, \city{M\u{a}gurele, Ilfov}, \postcode{077125}, \country{Romania}}}

\affil[3]{\orgname{Instituto de Astrof\'{i}sica de Canarias}, \orgaddress{\street{C/V\'{i}a L\'{a}ctea s/n}, \city{La Laguna}, \postcode{38205}, \state{Tenerife}, \country{Spain}}}

\affil[4]{\orgdiv{Departamento de Astrof\'{i}sica}, \orgname{Universidad de La Laguna}, \orgaddress{\street{Avda. Astrof\'{i}sico Francisco S\'{a}nchez s/n}, \city{La Laguna}, \postcode{38200}, \state{Tenerife}, \country{Spain}}}

\affil[5]{\orgname{European Space Agency (ESA), European Space Astronomy Centre (ESAC)}, \orgaddress{\street{Camino Bajo del Castillo~s/n}, \postcode{28692} \city{Villanueva de la Can\~{a}da}, \state{Madrid}, \country{Spain}}}

\affil[6]{\orgname{Department of Mechatronics, Optics and Mechanical Engineering Informatics, Faculty of Mechanical Engineering, Budapest University of Technology and Economics}, \orgaddress{\street{Muegyetem rkp. 3., H-1111}, \city{Budapest}, \country{Hungary}}}

\affil[7]{\orgname{Aurora Technology B.V.~for the European Space Agency (ESA)}, \orgdiv{European Space Astronomy Centre (ESAC)}, \orgaddress{\street{Camino Bajo del Castillo~s/n}, \postcode{28692} \city{Villanueva de la Can\~{a}da}, \state{Madrid}, \country{Spain}}}

\affil[8]{\orgname{HE Space Operations B.V. for ESA, European Space Research and Technology Centre}, \orgaddress{\city{Noordwijk}, \country{Netherlands}}}

\affil[9]{\orgname{cosine Remote Sensing BV}, \orgaddress{\city{Warmonderweg 14, 2171 AH Sassenheim}, \country{The Netherlands}}}

\affil[10]{\orgname{Institute of Space and Astronautical Science, Japan Aerospace Exploration Agency (JAXA)}, \orgaddress{\city{Sagamihara, 252-5210 Kanagawa}, \country{Japan}}}

\affil[11]{\orgname{Archaeology, Environmental Changes, and Geo-Chemistry, Vrije Universiteit Brussel}, \orgaddress{\city{Brussels}, \country{Belgium}}}

\affil[12]{\orgname{Royal Belgian Institute for Space Aeronomy}, \orgaddress{\street{Av. Circulaire 3, 1180}, \city{Brussels}, \country{Belgium}}}

\affil[13]{\orgname{Department of Earth and Planetary Science, The University of Tokyo}, \orgaddress{ \city{Bunkyo, Tokyo}, \country{Japan}}}

\affil[14]{\orgname{Dipartimento di Fisica e Astronomia, Padova University}, \orgaddress{ \street{Vicolo dell'Osservatorio 3}, \postcode{35122}, \city{Padova}, \country{Italia}}}

\affil[15]{\orgname{NASA Johnson Space Center}, \orgaddress{ \street{2101 NASA Parkway, Mail Code XI}, \city{Houston, TX 77058-3696}, \country{USA}}}


\affil[16]{\orgname{DLR Institute of Space Research}, \orgaddress{ \city{Berlin}, \country{Germany}}}

\affil[17]{\orgname{Observatoire de la C\^{o}te d'Azur, CNRS, Laboratoire Lagrange}, \orgaddress{ \city{Nice}, \country{France}}}


\abstract
{
The HyperScout-H (HS-H) instrument is one of the payloads aboard ESA's Hera spacecraft. Hera is a planetary defence mission that aims to provide a detailed characterization of the near-Earth binary asteroid (65803) Didymos-Dimorphos after the NASA/DART mission impact. HS-H is a versatile dual-use payload, functioning as a hyperspectral imager that captures both images and spectral data within the 0.65--0.95 $\mu$m wavelength range. The observations from this instrument will offer key insights regarding the composition of the two bodies Didymos and Dimorphos, space weathering effects, and the potential presence of exogenous material on these asteroids. Thanks to its wide field of view ($\approx 15.5^\circ \times 8.3^\circ$ in paraxial approximation), HS-H will be able to monitor the system's orbital dynamic and dust environment, while both components of this binary asteroid remain in the field of view for most of the asteroid phase of the mission.  These results also complement the data obtained from other instruments in characterizing the geomorphological units.

The data that will be obtained by HS-H will enable the creation of maps highlighting key spectral features, such as taxonomic classification, spectral slope, and band parameters. This article presents the pre-flight calibration of the instrument, outlines the science objectives, and discusses the expected investigations. The instrument's capabilities are demonstrated through laboratory observations of two meteorite samples and a dedicated software toolbox was developed specifically for processing the instrument's data.
}



\maketitle

\section{Introduction}
\label{intro}

Near-Earth asteroids (NEAs) are small bodies of the Solar System with a variety of eccentric orbits that bring them close to the orbit of the Earth (according to the definition, their perihelion is less than 1.3~au). They have short lifetimes ($\leq~10^7$~years) in this orbital state \citep{2002aste.book..409M}, so these objects originated elsewhere and were transported to their current orbits by a replenishment mechanism. The main asteroid belt, located between the orbits of Mars and Jupiter (2.1--5.2~au) is considered the principal source of NEAs, in particular the 3:1 mean-motion resonance with Jupiter, at 2.5 au, and the $\nu_6$ secular resonance with Jupiter and Saturn, at 2.1 au \citep{2002Icar..156..399B}. Collisions play a key role in the replenishment of NEA population: i) during a collisional family forming event, either by total disruption or by cratering-forming, numerous small fragments are produced; ii) transport mechanisms, such as the Yarkovsky effect \citep{2003Icar..163..120M} are more efficient for small objects than for big ones; iii) those small fragments have therefore their orbital semi-major axis more efficiently changed, reaching the unstable regions in the main belt (the 3:1 and the $\nu_6$ resonances), having their eccentricity increased and sent to the vicinity of our planet.

Consequently, NEAs can hold the records of the collisional history of the inner Solar System. Furthermore, they can shed light on the delivery of water and organic-rich material to the early Earth, and the subsequent emergence of life \citep{2013ApJ...767...54I, 2016E&PSL.441...91M}. Indeed, the history of our planet has been shaped by the impact with these small bodies \citep[e.g.][]{1980Sci...208.1095A}. The past recent impacts such as those at Chelyabinsk\citep{2013Natur.503..235B, 2013Natur.503..238B} and Tunguska\citep[e.g.][]{1993Natur.361...40C}, and the terrestrial impact cratering record \citep{1992Tectp.216....1G} represent clear evidences.

The proximity of NEAs at geological time scale not only makes them dangerous, but also among the best targets for robotic and human exploration \citep{2015aste.book..855A}. In this framework, the Asteroid Impact and Deflection Assessment (AIDA) concept was proposed \citep{2015AcAau.115..262C,2016AdSpR..57.2529M} as a demonstration and validation of the technology needed to deflect a hazardous asteroid by means of a kinetic impactor. Two missions were considered to perform and characterize the first controlled experiment of a large-scale hypervelocity kinetic impact. The first one, the NASA Double Asteroid Redirection Test -- DART \citep{2021PSJ.....2..173R, 2022PSJ.....3..244S, 2024PSJ.....5...49C} -- successfully impacted Dimorphos, the natural satellite of the binary near-Earth asteroid (65803) Didymos, on September 26, 2022 \citep{2023Natur.616..443D,2023Natur.616..457C, 2024Natur.627..505D}. 

The second one, the ESA Hera planetary defence mission, aims to provide detailed characterization of the (65803) Didymos system and to accurately measure the effects produced by the NASA/DART impact \citep{michel2022psj}. The mission takes advantage of a wide variety of instruments. The payload consists in two Asteroid Framing Cameras (AFCs), a hyperspectral imager (HyperScout-H), an altimeter, a Thermal Infrared Imager (TIRI), and a radio science experiment (RSE). Additionally, two cubesats will travel with the mission, Juventas and Milani  \citep{michel2022psj}. The observations performed by these instruments will allow a new understanding of the collisions in the small body population, of binary asteroids formation and evolution, of the asteroids' interiors, and of the effects of space-weathering on airless bodies.

ESA's Hera mission was successfully launched on October 7, 2024 from Kennedy Space Center, Cape Canaveral (Florida), and it will rendezvous with the asteroid in December 2026. Additionally, the ESA/Hera spacecraft swung by Mars in March 2025. During this orbital maneuver, the instruments performed observations of the planet and its natural satellites, Phobos and Deimos, targets of the JAXA Martian Moons eXploration (MMX) mission \citep{2022EP&S...74...12K}.

Near-Earth binary asteroid (65803) is a system with two objects, called Didymos and Dimorphos.  Its orbit has an eccentricity of $e = 0.383$ with an aphelion at 2.272~au, and a perihelion distance of 1.013~au, and it makes close approaches\footnote{\url{https://newton.spacedys.com/neodys/index.php?pc=1.1.0&n=65803}} to Earth and Mars at distances smaller than 0.05~au. Ground based observations \citep[e.g.][]{2022PSJ.....3..175P, 2024PSJ.....5...35M,2024PSJ.....5...17S, 2024Icar..41816138P}, together with the data collected by the DART mission camera and its cubesat \citep[LICIACube,][]{2023Natur.616..443D, 2024NatAs...8..445R,2024Natur.627..505D, 2023Natur.616..457C}, allowed to constrain its dynamical and physical properties pre and post impact.

The asteroid (65803) Didymos is classified as an S-type following its visible to near-infrared reflectance spectra, characterized by the presence of two absorption bands centred at 1~and 2~$\mu$m, and associated to mafic silicates \citep{2006AdSpR..37..178D, 2022PSJ.....3..183I,2023PSJ.....4..214R,2023PSJ.....4..229P}. S-types present an average spectrum resembling that of ordinary chondrites, the most common meteorites in our records. The size of Didymos was estimated to be $818 \times 796 \times 590$~m, with an uncertainty of $\pm 14$~m on all axes, while the size of Dimorphos is $173\pm1 \times 170\pm4 \times 113\pm1$~m \citep{2024PSJ.....5...49C, 2024NatCo..15.6202B, 2024PSJ.....5...24D, 2024PSJ.....5...74N}. \cite{2024NatAs...8..445R} reports a density of the bodies lower than 2\,400 kg/m$^3$ with a low fraction of boulders ($\leq 40$~\% of the volume). They infer that Dimorphos is a rubble pile that might have formed through rotational mass shedding and reaccumulation from Didymos. Interestingly, based on the ground-based spectral monitoring, \citet{2023PSJ.....4..229P}  suggests that Dimorphos was accumulated from weathered material ejected from Didymos’ surface. The smoothed particle hydrodynamics (SPH) impact simulations of \cite{2024NatAs...8..445R} indicate that the DART impact caused global deformation and resurfacing of Dimorphos. The NASA/DART mission decreased the orbital period of Dimorphos as a result of the kinetic impact with $33.0\pm1.0$~min \citep{2023Natur.616..448T, 2024PSJ.....5...35M}.

Spectroscopy is the most powerful remote sensing technique for determining the composition of celestial objects. The HyperScout-H (HS-H) instrument onboard the ESA/Hera spacecraft will provide spectral information of the observed targets. Hyperspectral imaging is a cutting-edge technology that enables efficient acquisition of spatial and spectral information at the same time. It gets the spectral profile at each spatial location in the field-of-view of the instrument \citep{Borengasser, Tsagkatakis}. Its use has been proven in a variety of remote sensing applications: climate change monitoring, medical diagnosis, precision agriculture, and food safety. Moreover, in the last years it has been used for satellite observations of the Earth. 

The HS-H instrument will provide hyperspectral images over the $\approx$0.650 - 0.950~$\mu$m wavelength range. The measurements obtained by the HS-H instrument represent a key element for understanding the surfaces of the Didymos-Dimorphos system: their composition, space-weathering effects, and the possible presence of exogenous material. The spectral properties of the material from the interior of Dimorphos, exposed by the NASA DART impact, will be revealed by HS-H. Furthermore, the images will complement the data obtained by the spacecraft's main cameras,  Asteroid Framing Cameras (AFC), which will be used for a detailed geo-morphological characterization of the system. 

This article presents an overview of the HS-H instrument, the laboratory calibrations, and its scientific objectives, as well as different methods to handle the observations. The paper is organized as follows: in Section \ref{section2} we describe the technical specifications of the instrument and the pre-flight calibration results; in Section \ref{section3}, we discuss the science objectives to be achieved using the data collected by HS-H. In Section \ref{section4}, we outline the methods to be used to process the data. As a proof of concept, in Section \ref{section5} we present the results obtained in the laboratory for two meteorite samples. The Conclusions are summarized in Section \ref{section6}.

\section{Instrument description}
\label{section2}

\begin{figure}
    \centering
    \includegraphics[width=1\linewidth]{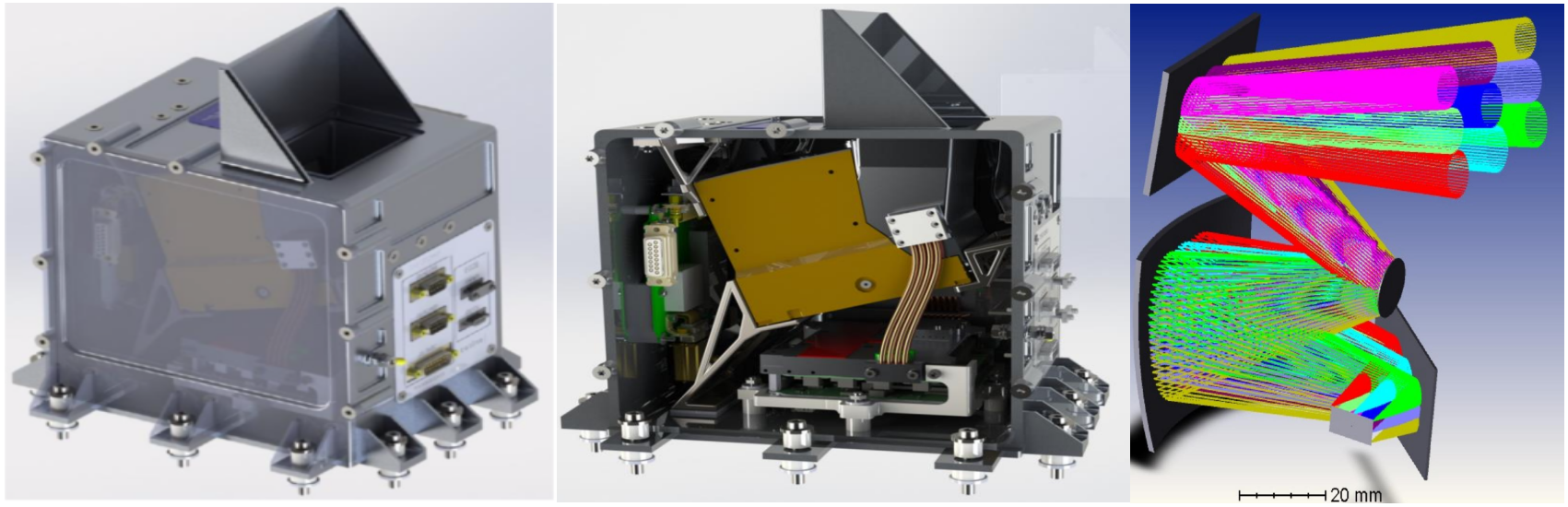}
    \caption{The left and central panels: rendered CAD assembly of the HS-H, without the removable plate and the multi-layer insulation blankets. Right: 3D optical layout of the telescope. Credit: \cite{hshiac}.}
    \label{fig:hshLayout}
\end{figure}

The HyperScout line of instruments \citep{2015MetroAeroSpace..547E, 2018AIAA....Esposito, 2019SPIE11180E..71E} has been developed by cosine Remote Sensing BV\footnote{\url{https://www.cosine.nl/}} and partners with the support from the European Space Agency (ESA) and the Netherlands Space Office. It was first developed as an Earth observation payload \citep{2024IJRS...45.2488B}. The HyperScout 2 platform, on which HS-H is built, is a versatile dual-use payload that provides both images and spectral information (multi-spectral imager). HS-H is a full-fledged, miniaturized hyperspectral instrument (Fig.~\ref{fig:hshLayout}) dedicated to planetary missions, providing operational efficiency and data volume optimization\citep{hshiac}. Below we present the instrument's technical specifications and report the results of the pre-flight calibration measurements.

\begin{table}[]
\caption{HS-H technical specification.}
\begin{tabular}{ll}
\hline
Parameter                & Value                                                                                                                   \\ \hline
Field of View            & \begin{tabular}[c]{@{}l@{}}$15.5^\circ~\times~8.3^\circ$ (paraxial)\\ $15.9^\circ~\times~9.9^\circ$ (real)\end{tabular} \\
Focal length             & 41.25 mm \\
Aperture diameter        & 10.3125 mm \\
Pixel pitch              & 5.5 $\mu$m \\
Number detector pixels   & \begin{tabular}[c]{@{}l@{}}$2048 \times 1088$\\ ($409 \times 217$ macropixels)\end{tabular}                            \\
Number of spectral bands & 25  \\
Central wavelengths [$\mu$m]      & 0.661, 0.670, 0.688, 0.702, 0.713,\\
 (median values)         & 0.730, 0.741, 0.754, 0.769, 0.782,\\
                         & 0.790, 0.806, 0.817, 0.830, 0.844,\\
                         & 0.855, 0.867, 0.881, 0.893, 0.901,\\
                         & 0.916, 0.924, 0.934, 0.944, 0.952 \\
Band FWHM  [$\mu$m]      & 0.0082, 0.0093, 0.0101, 0.0103, 0.0099, \\
 (median values)         & 0.0111, 0.0108, 0.0115, 0.0109, 0.0105, \\
                         & 0.0106, 0.0115, 0.0124, 0.0120, 0.0121, \\
                         & 0.0138, 0.0143, 0.0172, 0.0166, 0.0161, \\
                         & 0.0194, 0.0189, 0.0208, 0.0222, 0.0207, \\ 
Spatial resolution @ 1~km & \begin{tabular}[c]{@{}l@{}}13.2~cm (single pixel)\\ 66~cm (macropixel)\end{tabular}   \\ \hline
\end{tabular}
\label{hsh_spec}
\end{table}

\begin{figure}
    \centering
    \includegraphics[width=1.0\linewidth]{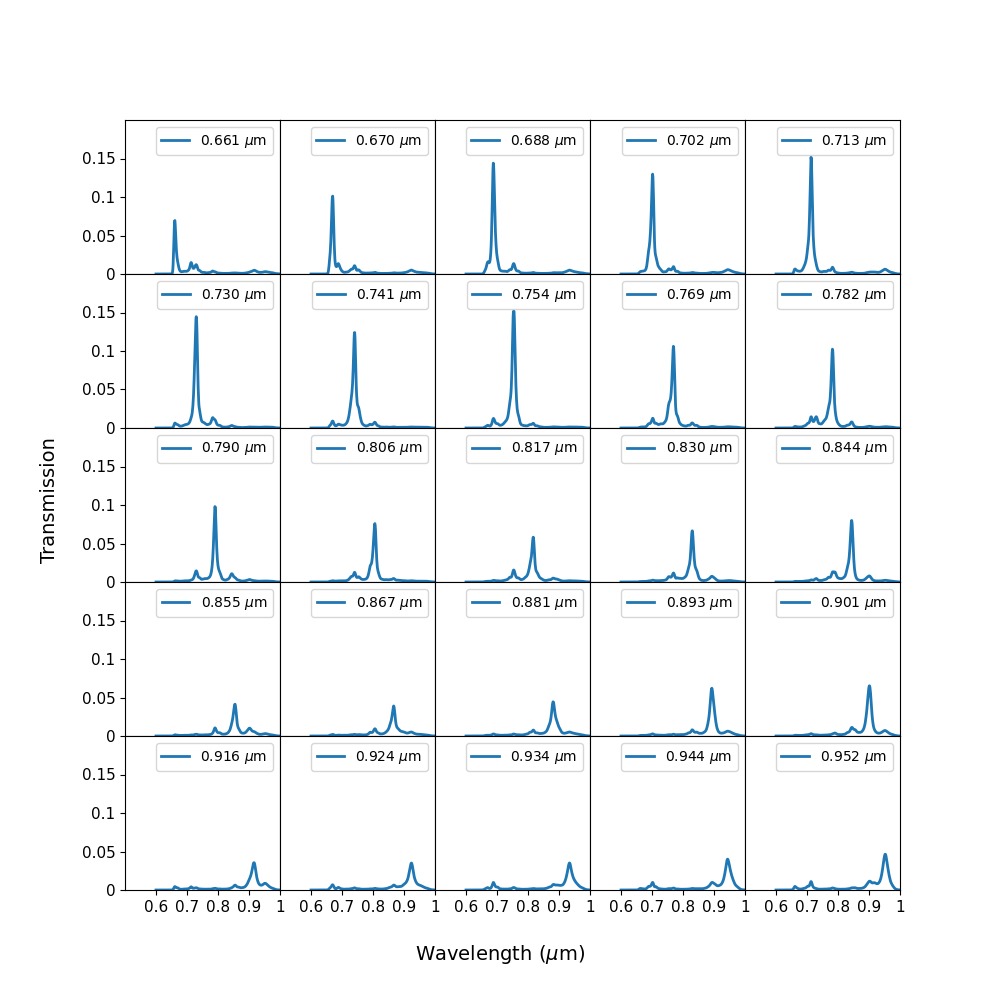}
    \caption{The HS-H macropixel configuration. Each subpixel represents a narrow band filter, with a specific central wavelength and an optical transfer function. The legend of each subplot shows the central wavelength (computed as the maximum of the optical transfer function), and the figure layout is the same as the macropixel one.}
    \label{fig:hsh_bands}
\end{figure}

\subsection{Technical specification}
\label{technicalspecification}

The HS-H has a mass of 5.145 kg and a volume of $238 \times 190 \times277$~mm$^3$. The main components are the telescope and the detector. A summary of the technical specifications of the instrument is provided in Table~\ref{hsh_spec}.  The telescope has an aperture of 10.3125~mm and a focal length of 41.25~mm. It is built using a Three-Mirror Anastigmat (TMA) configuration (Fig.~\ref{fig:hshLayout}), and it covers a wide field of view \citep{hshiac}. This is about $15.5^\circ \times 8.3^\circ$ in paraxial approximation, but because of distortions the real field of view covers $15.9^\circ \times 9.9^\circ$ (Table~\ref{hsh_spec}). 

The spectral window observed by HS-H spans approximately the wavelength range 0.65--0.95~$\mu$m. This is sampled by 25 narrow band filters arranged in a \emph{macropixel} structure, which is defined as a $5\times5$  matrix of pixels (to avoid any confusion, we will call them $subpixels$). Because the detector has 2048 columns and 1088 rows of subpixels, the format is translated into 410$\times$218 macropixels, with the last column and row containing incomplete macropixels. 

The detector is a Silicon Complementary Metal-Oxide-Semiconductor (CMOS) array coated with pixelated Fabry-P{\'e}rot interference filters \citep{hshiac}. The analog-to-digital converter (ADC) has 12 bits. The bias level and the gain of the amplifiers can be set using the software control interface. In this setup with the telescope, the average angular resolution is  0.132~mrad (27.25~arcsec). Spatially, from a distance of 1~km, the instrument will observe with an image scale of 13.2~cm/subpixel or 66~cm/macropixel.  Preliminary laboratory measurements show that the full width at half maximum of the point spread function varies between 1.2 and 2 subpixels. Nevertheless, this parameter will be characterized based on data obtained in-flight for various star fields.
 
The macropixel configuration follows the layout shown in Fig.~\ref{fig:hsh_bands}. The narrow band filters are ordered from left to right and from top to bottom in an ascending order (considered the top left corner of the image as a reference) of the central wavelength of the spectral band.  This macropixel pattern is repeated across the entire detector, starting with the top left corner. Fig.~\ref{fig:hsh_bands} also shows the optical transfer function (which includes the transfer function of the Fabry-P{\'e}rot interference filters and the quantum efficiency of the CMOS detector) and the central wavelength (defined as the position of the maximum of the optical transfer function) for each narrow band filter.  Throughout the paper, we will refer to all subpixels with the same narrow band filter as a wavelength channel.

The median central wavelengths (defined as the position of the maximum of the filter response) and the median Full Width at Half Maximum (FWHM) corresponding to each channel are shown in Table~\ref{fig:hsh_bands}.  These values were reported in the test report provided by the instrument producer, \emph{cosine Remote Sensing BV}.  Minor variations exist in filter profiles across the detector, with an average central wavelength deviation of $\pm1$\%. 

The FWHM of individual bands ranges from 0.008 to 0.022 $\mu$m. The filters’ transfer functions exhibit ripples; therefore, the \emph{In-Band (IB)} spectral region is defined as twice the FWHM of the peak detection, while the remainder of the spectral range is designated as \emph{Out-of-Band (OOB)}. The average ratio between IB and OOB is 0.7.  These ripples must be taken into account when comparing HS-H spectra with data from other spectrographs, as a conversion using the HS-H filter responses is required before comparison.

Spatial stray light is defined as light that reaches the detector through unwanted optical paths. To mitigate this effect, the HS-H manufacturer designed a complex system of baffles to prevent light from reaching the detector without being reflected by all the telescope mirrors. The instrument is a fully reflective telescope with four metallic mirrors. Therefore, no unwanted specular reflections (ghosts) can reach the detector and degrade the image quality. Based on these considerations, the effect of stray light can be neglected. This has been demonstrated by the manufacturer through analysis and testing as well as in orbit with the previous version of the instrument, which had the same telescope design as the present HS-H.

\subsection{Pre-flight calibration}
Two independent sets of calibration data for HS-H were obtained. The first one was obtained by the instrument manufacturer, \emph{cosine Remote Sensing BV} and provided alongside with test report to ESA. The second one was obtained at the laboratories of the European Space Research and Technology Centre (ESTEC) of ESA. Any disagreements between the results were reviewed and could be resolved, so both reports are consistent. In this section, we outline the results of the optical calibration done at ESTEC. The obtained calibration files, i.e., master-flat and master-dark, will be used for processing the data that will be obtained during the mission.

The calibration equipment included a Spectralon 99\% diffuse target, which is a calibrated and certified reflectance standard (SRS-99-020), and a halogen lamp, Mikropack HL-2000-FHSA-HP (Serial No. 034990084). The HS-H was mounted on the optical bench with the detector aligned horizontally. Calibration tests were conducted using the default settings for the gain and the offset of the analog digital converter (ADC) block, which are the same as those planned for the mission. For most of the measurements, full-frame images (2048$\times$1088 subpixels) were acquired. The camera also supports sub-frame acquisition (also called \emph{region of interest} or ROI), though this mode was only used in very specific cases. Exposure times were set between 0.1~ms and 1800.0~ms, depending on the particular calibration measurement. The obtained calibration images can be accessed through the MOGI webserver\footnote{\url{https://didymos.mogi.bme.hu/}} which also provides access to the HS-H images.

The 41.25~mm focal length of the HS-H optical system allows for relatively close defocused imaging at about 10 m. Long-distance, in-focus imaging is not an issue since the optical system is designed for both vacuum and infinity focus. While the slight defocusing of the ground images is not a problem for flat and radiometric measurements, it does make ground distortion analysis more difficult and less accurate. Therefore, the final distortion correction will be based on in-flight images of star fields.

Nevertheless, the space environment can degrade the performance of the HS-H. Therefore, we plan to acquire calibration images every six months during the cruise phase and at each asteroid mission phase to monitor the detector's behavior. These data will include bias and dark frames, star fields, and standard star observations. 

In addition, on March 12, 2025, the ESA/Hera mission performed a Mars flyby. During this event, images of Mars and its satellites were acquired by AFC, HS-H, and TIRI. These images will be used to cross-validate the functionality of the instruments and to compare the data acquired by Hera with that obtained by previous Martian missions.

\subsubsection{Dark Current}

\begin{figure*}
    \centering
    \includegraphics[width=0.5\linewidth]{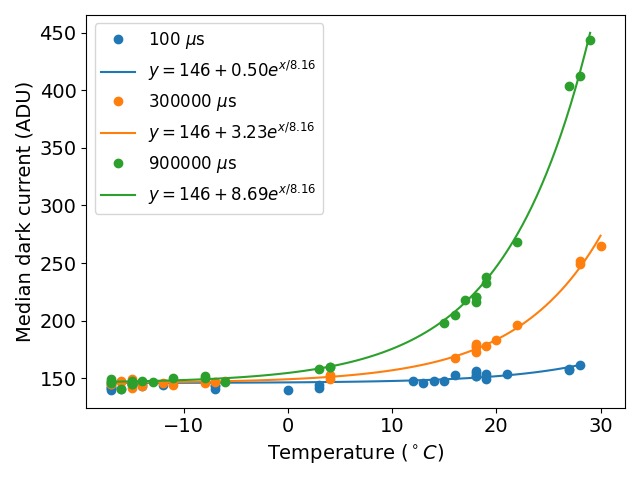}
    \includegraphics[width=0.5\linewidth]{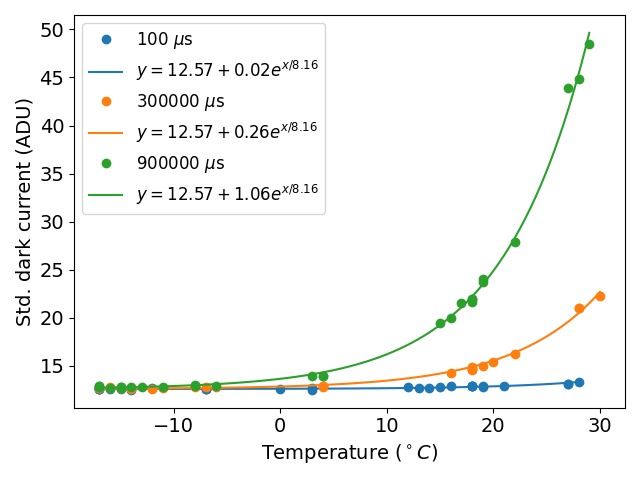}
    \caption{The variation in the median and standard deviation of pixel values of dark current as a function of temperature. Both plots are shown for three different exposure times, with the behavior approximated using an exponential function. }
    \label{fig:darks}
\end{figure*}

The dark current is defined as the signal received from the device when it is not illuminated by
light. The dark images were captured at ambient room temperature and within the Thermal Vacuum Chamber, using both the Large Vacuum Facility (LAVAF) and the Little Vacuum Facility (LIVAF).  The characteristic curves are provided for three exposure times, 0.1~ms, 300~ms, and 900~ms (Fig.~\ref{fig:darks}). The maximum exposure setting allowed by the instrument is 9.99~sec, while the minimum exposure time is 0.1~ms. This represents the closest approximation for the bias level of the detector and it has a temperature dependency in the range of $\approx10$~ADU. 

HS-H has eight temperature sensors which are read by the Instrument Control Unit (ICU). The temperature reported in the image header (with the software version available when the tests were performed) corresponds to the detector sensor, and it is used to trace the characteristic curves.  With the current version of the software, the precision of the temperature reported value is $\pm 1^\circ$ (the sensor value is read as an integer). During the ambient calibration tests, the sensor-reported temperature was between $12.0^\circ$C and $18.0^\circ$C. The laboratory temperature was between $\approx 20^\circ$C and $\approx 24.5^\circ$C. The offset between the temperature reported by this sensor and that near the Focal Plane Array (FPA), indicated by the instrument manufacturer as the reference, also confirms the temperature difference of $\Delta T = 7.8 \pm 1.7^\circ$C. This offset was computed later based on in-flight data, as the initial software version did not report FPA sensor temperatures.

\begin{figure*}
    \centering
    \includegraphics[width=0.5\linewidth]{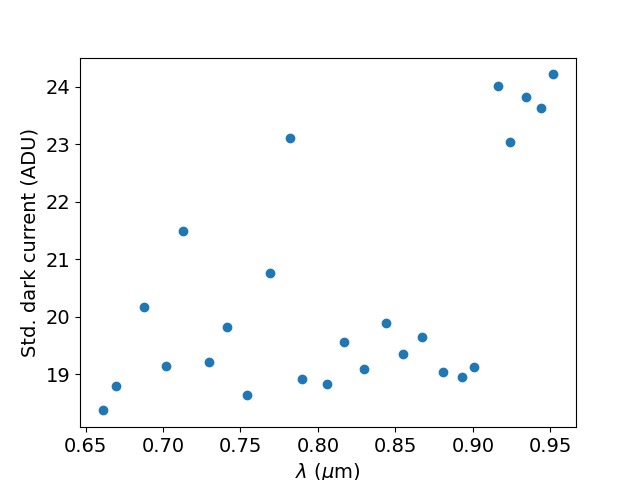}    
    \caption{The variation in standard deviation of dark current for each narrow band filter. (represented by the central wavelength of each filter). The curve is plotted at a temperature of 18$^\circ$C (as indicated by the internal sensor) and an exposure time of 0.9 s.}
    \label{fig:stddarks}
\end{figure*}

To monitor the behavior, we computed the median $m$ and the standard deviation $\sigma$ of each exposure according to
\begin{eqnarray}
m(t_{\mathrm{exp}},T)[\mathrm{DN}] &=& 146+(9.10 \, t_{\mathrm{exp}}[\mathrm{s}] + 0.5 ) \cdot \mathrm{e}^{0.1225 \, T[^\circ\mathrm{C}]} ,
\label{median_level} \\
\sigma_{\mathrm{noise}}(t_{\mathrm{exp}},T)[\mathrm{DN}] &=& 12.57+(1.18 \, t_{\mathrm{exp}}[\mathrm{s}] - 0.0252) \cdot \mathrm{e}^{0.1225 \, T[^\circ\mathrm{C}]} .
\label{sigma_level}
\end{eqnarray}
The variations are shown in Fig.~\ref{fig:darks}. 

The dark current shows an exponential dependency on temperature \citep{2008eiad.book.....M}. It can be expressed in the form $n_D(T) = n_D(T_0) \, \exp\left(\frac{T-T_0}{\alpha}\right)$, where $T_0$ is a reference temperature and  $\alpha$ is the temperature deviation.  Following these approaches, the variation of median level $m$ and standard deviation $\sigma$ of all pixel values is modeled with an exponential function of the form $c_0+(c_2 \, t_{\mathrm{exp}} +c_3 ) \cdot \exp({c_4 \, T})$.  The equations \ref{median_level} and \ref{sigma_level} provide the coefficients of these functions and allow to estimate the median pixels level $m(t_{\mathrm{exp}},T)$ and the pixels values standard deviation (the noise level) $\sigma_{\mathrm{noise}}(t_{\mathrm{exp}},T)$ for any exposure time $t_{\mathrm{exp}}$ and temperature $T$.

We found that the median level remains constant across all channels (the same value is recorded for all narrow band filters), while the standard deviation $\sigma_\mathrm{noise}$ varies between channels, with the longest wavelengths being the noisiest (Fig.~\ref{fig:stddarks}).  This assessment was made with images obtained at $18^\circ$C temperature and exposure times of 0.9~s.

\begin{figure}
    \centering
    \includegraphics[width=1\linewidth]{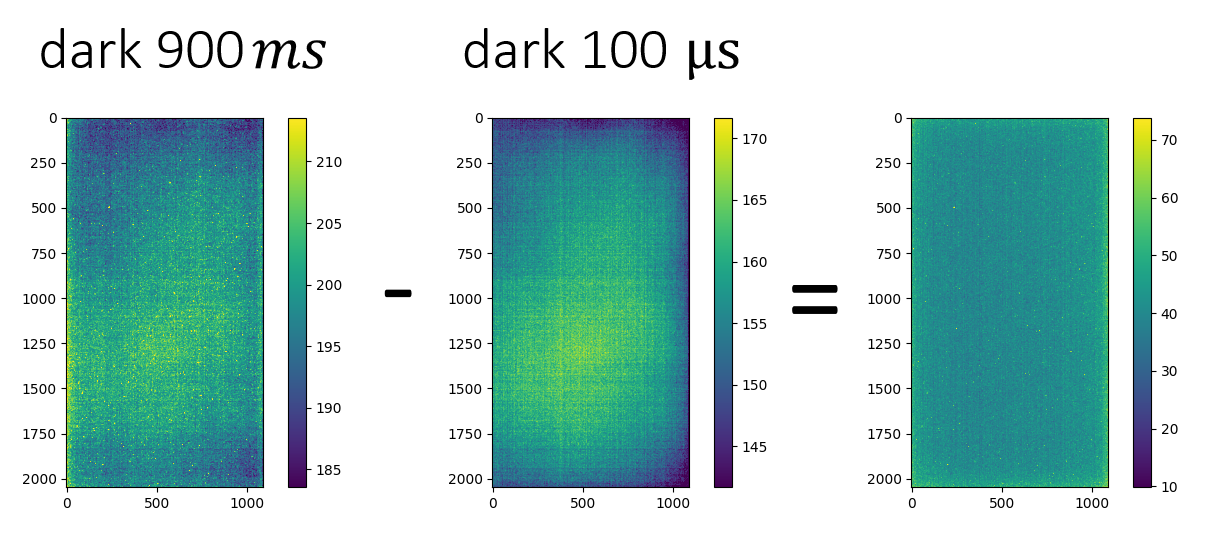}
    \caption{ The comparison between the median master dark obtained for an exposure time of 900~ms and the one for 0.1~ms. }
    \label{fig:masterdark}
\end{figure}

This dataset was used to generate the master dark frames (obtained as the median of all images taken for a given exposure time and sensor temperature). Two examples of the master darks, along with their difference, are shown in Fig.~\ref{fig:masterdark}.  Since 0.1~ms is the minimum allowed exposure time, the central image from Fig.~\ref{fig:masterdark} serves as the closest approximation to the bias. The difference between the left image (900~ms exposure) and the central image (0.1~ms) reveals the spatial structure of the dark current, which does not exhibit any discernible pattern.

A faint pattern can be observed across the detector, which remains consistent across varying exposure times. However, slight random variations in this pattern may occur with changes in temperature and the number of consecutive exposures, suggesting a dependence on the temperature distribution across the detector. This effect only became evident when comparing the 0.1~ms master dark obtained on Earth with that acquired in space, with the estimated variation being approximately $\sigma \approx 4$~DN, less than a third of the level of the camera's pixel-to-pixel noise. This behavior will continue to be monitored using data collected in space.

For data obtained in space, correction for bias and dark levels is performed by subtracting from the raw image the master dark at 0.1 ms exposure (serving as the bias equivalent) and a constant offset $b$, which is computed as the median of the background.

\subsubsection{Bad subpixels map}

Bad pixels are regions on a CMOS detector that produce invalid digital number (DN) values. These can be single pixels, lines, rows, or larger areas. The current bad pixel map used in the calibration pipeline is generated from a dark frame acquired at $19^\circ$C. This will be updated based on the calibration data acquired during the flight. Individual subpixels with signals falling outside a $\pm~5\sigma$ range are flagged as bad, where $\sigma$ is the standard deviation for that specific channel.

Initial analysis has identified a small number of bad pixels. This includes a malfunction in the first two rows and the final row of the detector. Only 12 additional individual hot pixels have been found across the entire detector array.

Individual bad pixels are corrected using a local interpolation method. The value of each bad pixel is replaced by the mean of its valid, neighboring pixels within the same channel:
\begin{equation}
\mathrm{ImgCorrect}(i,j) = \frac{1}{4} \left( \mathrm{Img}_{i+5, j+5} + \mathrm{Img}_{i-5, j+5} + \mathrm{Img}_{i+5, j-5} + \mathrm{Img}_{i-5, j-5} \right)
\label{badpix}
\end{equation}
The subpixel at coordinates (i,j) in the corrected image (ImgCorrect) is replaced by the average of four neighboring subpixels from the same channel. The neighbors are located five pixels away in each cardinal direction (left, right, top, and bottom). However, the three malfunctioning rows are replaced with NaN values in the calibrated images.

\subsubsection{Flat-field correction}

To acquire the images required for flat-field calibration, we used a 0.5~m diameter integrating sphere and a halogen lamp with a continuous spectrum covering the HS-H wavelengths. The camera was mounted horizontally, with its optical axis aligned toward the entrance aperture of the integrating sphere. The sphere was positioned close to the camera to avoid vignetting from the aperture. The back surface of the sphere was illuminated by halogen lamps and the current was gradually increased to 1.6~A. The lamp's power-up sequence is initiated by a current ramp-up phase lasting 5 minutes. This controlled increase prevents inrush current from damaging the lamp filament. Throughout the subsequent operational period, the current is actively regulated and maintained at a constant 1.6~A with a stability of $\pm0.001$~A. During this measurement phase, the lamp's operating voltage was observed to be highly stable, with a maximum variation of less than 0.005~V. After 10~minutes (warm-up time according to the lamp data sheet), the lamps stabilized, and the images were taken using various exposure times.

First, we selected all images with a median level between 1300 and 2200~ADU, corresponding to exposure times ranging from 3.5 to 5.5~ms. This selection avoids saturation in any channel and ensures a sufficient signal-to-noise ratio (SNR). Dark subtraction was then applied to all of the selected images.

\begin{figure}
    \centering
    \includegraphics[width=0.5\linewidth]{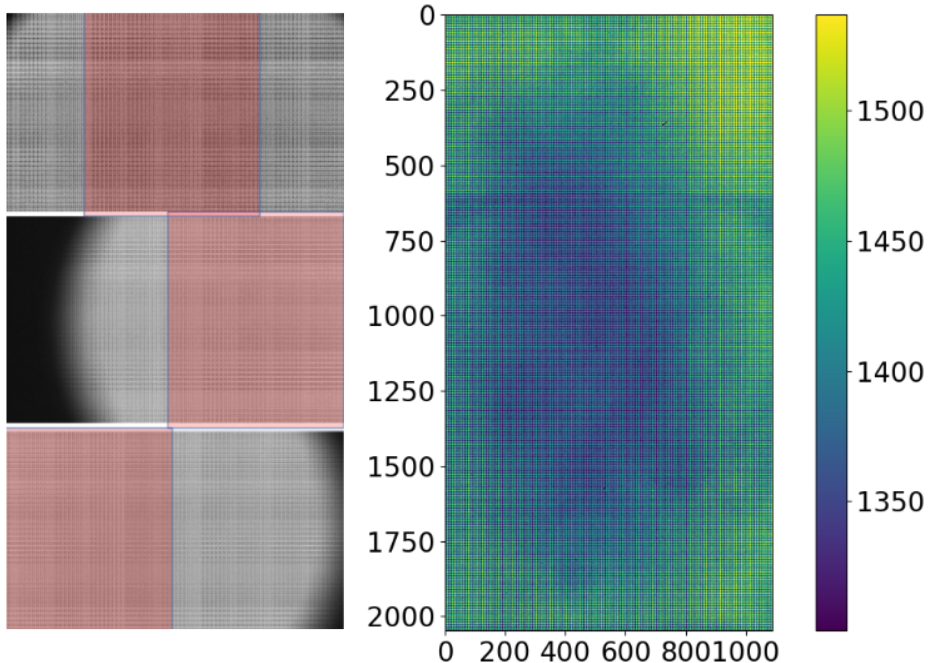}
    \caption{Left - the merging of different images to obtain the master-flat. Right - the master-flat obtained after combining and merging all images, as explained in the text.}
    \label{fig:master-flat}
\end{figure}

The illumination uniformity is primarily a result of the integrating sphere. Before the measurement, this was verified using a highly sensitive 16-bit reference camera. This test revealed an illumination non-uniformity of less than 0.05\%. Nevertheless, the integrating sphere did not provide uniform illumination across the entire detector; thus, it was slightly repositioned between different exposures. As a result, the images were divided into three sets, one with the left side illuminated uniformly (5 images), another with the center part (7 images), and the last with the right side (5 images). To work with only the uniform-illuminated regions, we cropped the images, keeping only the relevant sections, resulting in subframes of $1088 \times 800$ pixels (Fig.~\ref{fig:master-flat}).

The next step was to normalize each trimmed image to its median value. Then, we combined the images within each subset (left, center, and right) using the median method. Finally, to create the flat-field, we merged the three obtained median subframes corresponding to each subset, aligning them using the overlapping regions. This resulting image still contains the spectrum of the broadband lamp. To remove it, we first converted the normalized broadband spectrum of the halogen lamp using the HS-H filters transfer function (Fig.~\ref{fig:hsh_bands}):
\begin{equation}
S^i_{\mathrm{HS-H}} = \frac{\int{F^i(\lambda) \, S(\lambda) \, \lambda \, \mathrm{d}\lambda}}{\int{F^i(\lambda)\, \lambda \, \mathrm{d}\lambda}}
\label{eqhalogenlampsspectrum}
\end{equation}
This spectral conversion follows Eq.~\ref{eqhalogenlampsspectrum}, where $S^i_{HS-H}$ is the value for the converted spectrum for filter $i$,  $F^i(\lambda)$ is the transfer function for channel $i$, and $S(\lambda)$ is the spectrum to be converted.

\begin{figure*}
    \centering
    \includegraphics[width=0.95\linewidth]{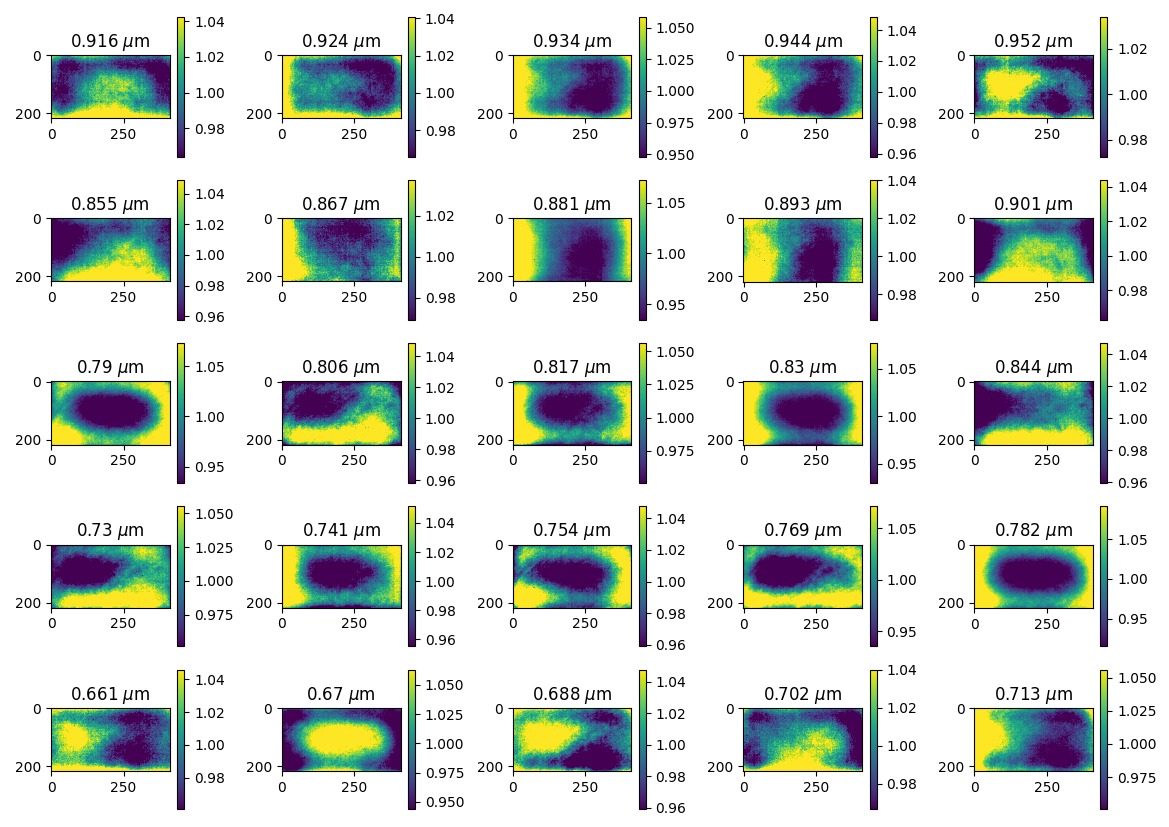}
    \caption{The master flat for each wavelength channel. Each image was normalized so that its average value is 1. }
    \label{fig:master-flat-channels}
\end{figure*}

Finally, we quantified the pattern variations across each channel, as shown in Fig.~\ref{fig:master-flat-channels}. We obtained the 25 flats by demosaicing the image. Each of them was normalized to the median value.  One can notice that the pattern is different from one channel to the other, and the variations are of the order of 10\%. All the channels are normalized to the average value. Flat-field correction is performed by dividing the science image by the master flat image.

The flat field presented here is consistent with the one obtained from the test data provided by the instrument manufacturer, who performed the measurements using different instruments and a different setup. The observed difference is less than $\pm 2$\% and was subsequently attributed to the illumination conditions in the manufacturer's setup. As a result, both master flats derived in the laboratory are considered equivalent. A definitive validation of the laboratory-derived flats can only be performed by using the images acquired during the Mars swing-by, when the Martian surface covers the entire instrument's field of view in four separate exposures.

\subsubsection{Linearity test and radiometric calibration}

The radiometric measurements are a set of tests designed to quantify how the detector responds to the incident light. The aim is to obtain a predictable relationship between the light intensity and the DN recorded by each subpixel.  The same broadband lamp as for the flat-field acquisition was used.

\begin{figure}
    \centering
    \includegraphics[width=0.95\linewidth]{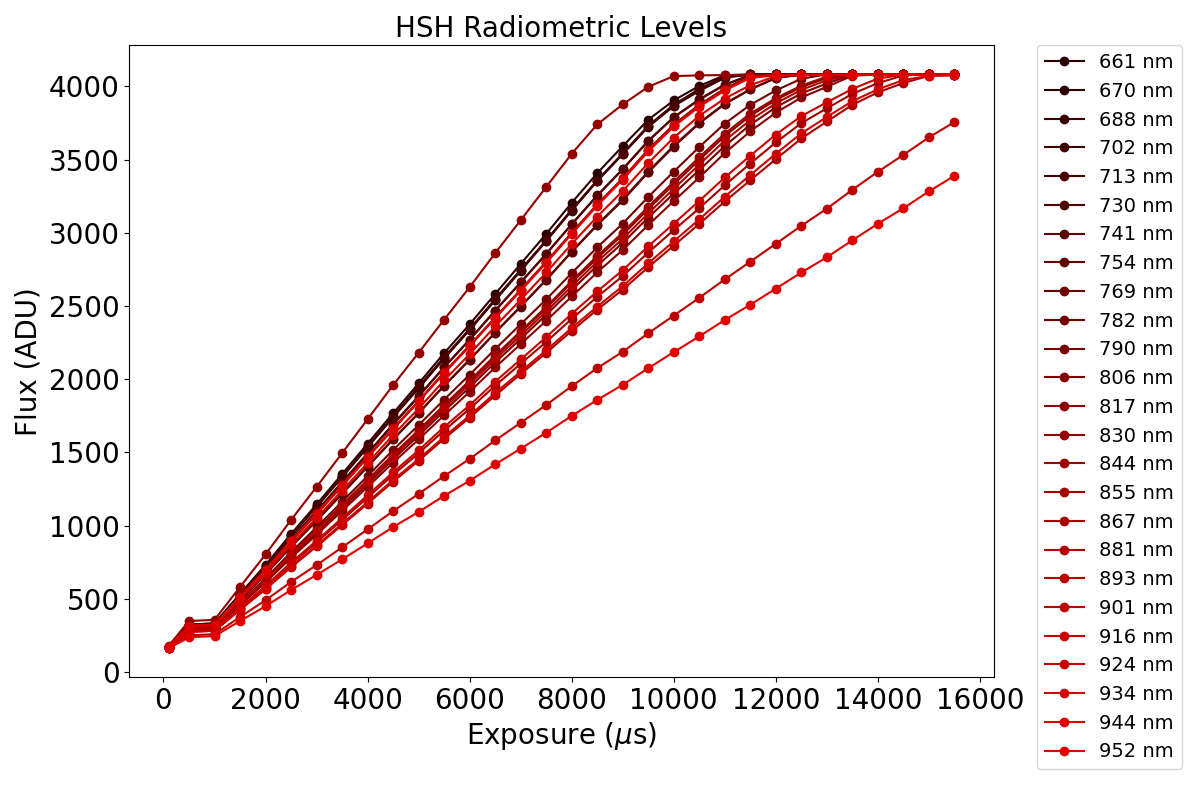}
    \caption{The results of linearity testing for each channel of the HS-H detector.}
    \label{fig:radiometric_wv}
\end{figure}

The linearity test result shows that the detector's output signal is directly proportional to the exposure time when the intensity of the lamp is kept constant. A perfect linearity is obtained between $\approx 300$ and $\approx 3800$~DN. The results are shown in Fig.~\ref{fig:radiometric_wv}.

To obtain the constants that allow us to convert between the DNs and light intensity units, we used the same images as for the flat. These constant decouple the response from the flat-field. The sensitivity of the detector was computed as the ratio between the average recorded signal in DN, $N_\mathrm{DN}$, on each channel, scaled with the exposure time -- $\frac{N_\mathrm{DN}}{\Delta t}$ -- and the radiance of the integrating sphere. The uncertainties ($\sigma_\mathrm{Sensitivity}$) were computed as the standard deviation of the signal recorded by the pixels corresponding to the same channel. They do not take into account the stability of the light source output which is 0.25\% peak-to-peak according to the manufacturer's specifications.  The results are shown in Table~\ref{hsh_calib}. The relative gain, representing the variation in response across different channels, is reported relative to the 0.817~$\mu$m channel, for which we assume a gain of~1.

\begin{table}[]
\caption{HS-H sensitivity factors. The average pixel values in DN, the source radiance, the sensitivity factors, their errors, and the relative gains are shown for each channel identified by central wavelength $\lambda$.}
\begin{tabular}{c c c c c c}
\hline
$\lambda$ & Avg. record & Source radiance & Sensitivity       & $\sigma_{Sensitivity}$ & Relativ gain \\
$[\mu$m]  & [DN]        & [W/m2/sr/nm]    & [DN/(J/m2/sr/nm)]  & [DN/(J/m2/sr/nm)] & \\ \hline
0.661 & 820.12  & 0.0583846 & 3511733.65 & 1975.35 & 0.64 \\
0.670 & 940.67  & 0.0568248 & 4138484.05 & 2327.89 & 0.73 \\
0.688 & 1302.83 & 0.0579805 & 5617516.94 & 3159.85 & 1.01 \\
0.702 & 1361.84 & 0.0598579 & 5687786.40 & 3199.37 & 1.06 \\
0.713 & 1557.02 & 0.0609827 & 6383036.07 & 3590.45 & 1.21 \\
0.730 & 1553.51 & 0.0631121 & 6153758.34 & 3461.48 & 1.21 \\
0.741 & 1512.49 & 0.0638740 & 5919827.94 & 3329.90 & 1.18 \\
0.754 & 1813.71 & 0.0652187 & 6952424.99 & 3910.73 & 1.41 \\
0.769 & 1542.25 & 0.0664358 & 5803528.41 & 3264.48 & 1.20 \\
0.782 & 1588.89 & 0.0672072 & 5910430.22 & 3324.61 & 1.24 \\
0.790 & 1474.00 & 0.0685706 & 5374010.65 & 3022.88 & 1.15 \\
0.806 & 1344.48 & 0.0689285 & 4876361.74 & 2742.95 & 1.05 \\
0.817 & 1283.87 & 0.0692935 & 4631988.38 & 2605.49 & 1.00 \\
0.830 & 1436.80 & 0.0700076 & 5130865.69 & 2886.11 & 1.12 \\
0.844 & 1560.38 & 0.0702493 & 5553006.64 & 3123.56 & 1.22 \\
0.855 & 1174.67 & 0.0697290 & 4211542.71 & 2368.99 & 0.91 \\
0.867 & 1138.34 & 0.0691802 & 4113691.13 & 2313.95 & 0.89 \\
0.881 & 1315.46 & 0.0691326 & 4757017.49 & 2675.82 & 1.02 \\
0.893 & 1510.30 & 0.0693202 & 5446834.39 & 3063.84 & 1.18 \\
0.901 & 1604.68 & 0.0692521 & 5792903.76 & 3258.50 & 1.25 \\
0.916 & 1210.78 & 0.0683875 & 4426182.84 & 2489.72 & 0.94 \\
0.924 & 1177.08 & 0.0682012 & 4314735.62 & 2427.03 & 0.92 \\
0.934 & 1272.53 & 0.0681067 & 4671072.35 & 2627.47 & 0.99 \\
0.944 & 1398.53 & 0.0682791 & 5120651.26 & 2880.36 & 1.09 \\
0.952 & 1554.70 & 0.0682911 & 5691450.69 & 3201.44 & 1.21 \\ \hline
\end{tabular}
\label{hsh_calib}
\end{table}

\begin{figure*}
    \centering
    \includegraphics[width=0.9\linewidth]{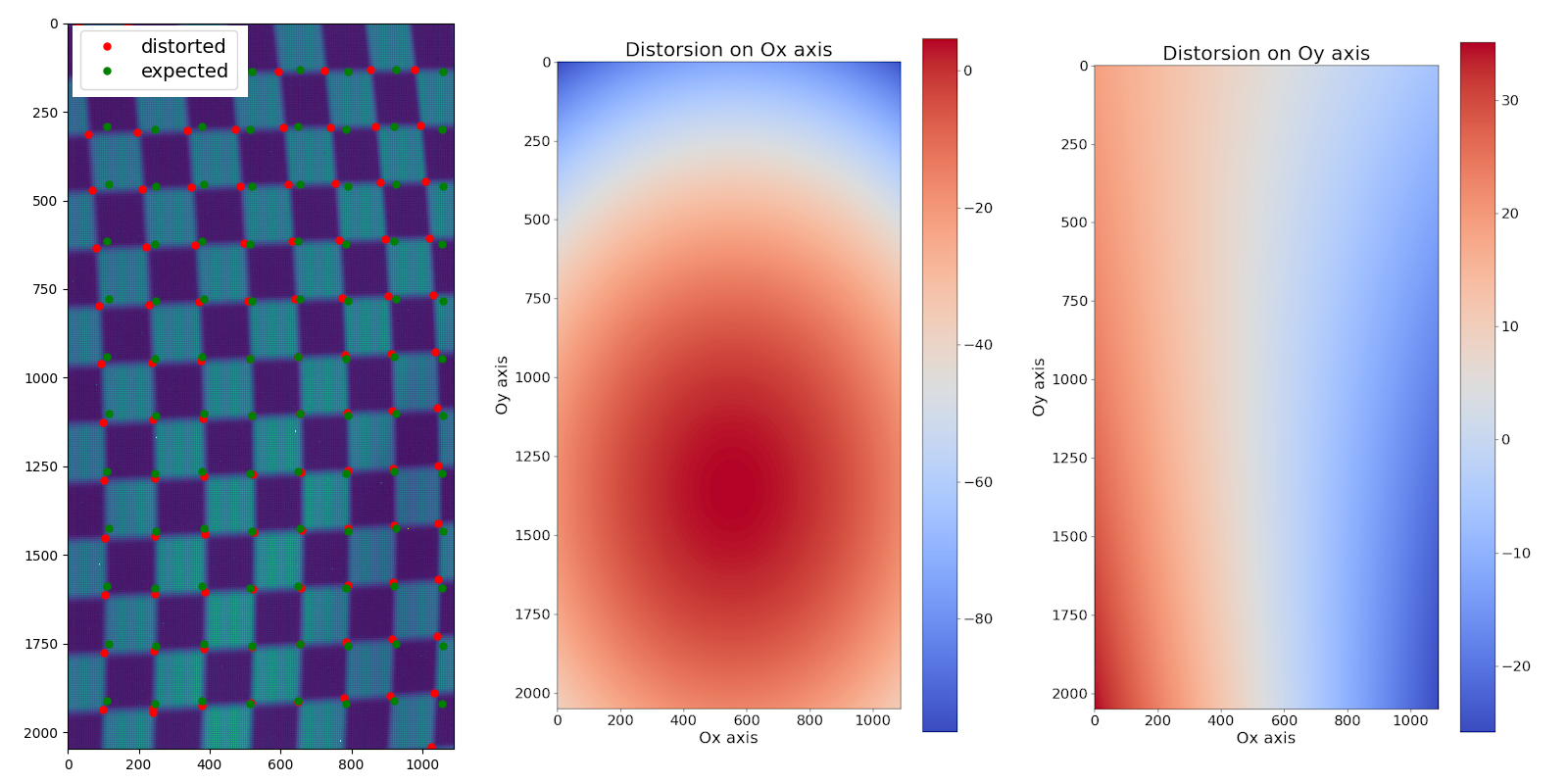}
    \caption{First left side image shows the positions of the corners that are subject to geometrical distortion (red points) and the positions (green points) that one would expect in the absence of distortion. The other two images are the distortion heatmaps showing the difference between the expected and real values of the coordinates (in pixels) derived by two-dimensional curve fitting on each coordinate of Ox and Oy axes.}
    \label{fig:distorsion}
\end{figure*}

\subsubsection{Distortion corrections}
To quantify the geometrical distortions, we targeted a checkerboard (a board on which white and black squares were printed), as shown in Fig.~\ref{fig:distorsion}. The distortion is quantified by determining how much the coordinates of each pixel are shifted with respect to the expected position. 

To evaluate the required distortion corrections, we use the corners of each checkerboard square. Firstly, we identify the corners in the HS-H image. This was done by identifying the transitions between the black and white squares, and then finding the intersection points between the horizontal and vertical curves. Then, we compare their position with the expected ones obtained by extrapolating the lines corresponding to the middle squares.

As a first approach, using these pairs of points, we determine the shifts and we fit functions $f_x(x,y)$ and 
respectively,
by curve fitting.
These functions provide the geometrical distortion corrected position for the subpixel observed at the $(x,y)$ image coordinates. The fitted functions we
obtain are
\begin{eqnarray}
f_y(x,y) &=& -95.0432033109 + 0.0811860901 \cdot y + 0.1099318927 \cdot x \nonumber
              \\
         & & {} - 0.0000732383 \cdot y^2 - 0.0000379289 \cdot x^2 - 0.0000000218 \cdot x  y ,
\label{fyd} \\
    f_x(x,y) &=& 18.8843072035 - 0.0259273937 \cdot y + 0.0031160476 \cdot x \nonumber
                \\
                & & {} + 0.0000024368 \cdot y^2
     + 0.0000023511 \cdot x^2 - 0.0000160065 \cdot x  y .
\label{fxd}
\end{eqnarray}
The $Ox$ axis is the longest dimension, and the $Oy$ axis is the shortest dimension of the camera, thus,
the $x$ values range from 0 to 2047 and the $y$ values from 0 to 1087.

These functions derived above will serve as an initial approximation for a precise calibration, which will be made using star field observations. As noted by \citep{2018Icar..300..341S}, the camera distortion parameters after launch may have changed from the pre-flight values due to severe vibration during the launch and large changes in environmental conditions. They used a fit of the form $r_u = r_d + k_3 \, r_d^3 + k_5 \, r_d^5$, where $r_d$ is the distorted radius and $r_u$ is the undistorted radius. Nevertheless,  the distortion distribution shown in Fig.~\ref{fig:distorsion} for HS-H is not symmetrical to the center of the field. This adds an additional constraint.



\section{Science Objectives}
\label{section3}

The Hera mission aims to obtain a detailed characterization of the Didymos-Dimorphos system in the context of the impact experiment performed by the NASA/DART spacecraft \citep{michel2022psj}. The determined properties will not only allow us to understand the physical phenomena generated by the large-scale impact, but it will also offer an unprecedented understanding of an S-type binary near-Earth asteroid. This compositional class is the most representative of the potentially hazardous asteroids that may pose a threat to the Earth in the future, and the most common class among the near-Earth asteroid population \citep[e.g][]{2010A&A...517A..23D, 2019A&A...627A.124P, 2019Icar..324...41B, 2023MNRAS.519.1677M}.

The HS-H instrument plays a key role in achieving the objectives of the ESA/Hera mission thanks to its unique capabilities. As it is a hyperspectral imager with the widest field of view (157~square degree, taking the distortions into account) among all the instruments onboard the spacecraft, it will allow linking the compositional information with the surface morphology on both asteroids, Didymos and its satellite, Dimorphos. Moreover, the environment around the Didymos-Dimorphos system will be scrutinized for the presence of boulders and dust ejected due to the impact, thanks to the HS-H wide field of view.

Another key topic for HS-H observations is the space weathering on airless bodies. This is the gradual alteration of materials when they are exposed to a variety of natural processes that occur in the space environment, such as the micro-meteorite bombardment or the impact from highly-charged solar wind particles \citep{2002aste.book..585C, 2016JGRE..121.1865P}. Using ground-based observations, the space weathering degree among the S-complex asteroids has been invoked to explain the spectral differences between the S-types, which are considered space-weathered surfaces, the Sq-types as intermediate space weathered surfaces, and the Q-type, which are interpreted as having fresh ordinary-chondrite-like compositions \citep[e.g.][]{2002aste.book..585C,2010Natur.463..331B,2024M&PS...59.1329M}. This interpretation is supported by laboratory experiments: spectral reddening and darkening, as well as a decrease in the absorption band depths, are changes observed in all experiments simulating space-weathering effects on silicates and meteorites \citep[][and references therein]{2015aste.book..597B}. These experiments also demonstrate that space weathering does not alter the band centers or the band area ratio. Interestingly, \citet{2012MNRAS.421....2M} observed that one of the effects of space weathering of silicate-rich asteroids, i.e., spectral reddening, results from a combination of three factors: exposure to the space environment over the object's lifetime, its original composition, and the structure and texture of its surface. 

Models of binary formation predict that Dimorphos was part of Didymos before the binary pair was separated, and thus they are assumed to represent the same composition \citep{2000Icar..148...12P,2008Natur.454..188W}. This assumption can be tested by measuring the spectral properties of the two bodies and quantifying any differences found. In addition, the impact crater -- or the reshaped part left over by the impact \citep{2024NatAs...8..445R} -- will provide the opportunity to study the composition of the material underneath the surface of the asteroid. We can assume that the material from the fresh DART impact crater (or the reshaped surface) will be unweathered and it is therefore expected to be brighter and bluer than the overall surface material of Didymos and Dimorphos. Thus, by mapping the space weathering degree across the two surfaces, we can characterize the re-deposition of the dust generated by the impact on both bodies. 

Furthermore, the ejecta thrown out by the impact have contributed to Dimorphos' velocity change, and this is considered an important effect for the understanding of the momentum transfer. However, the ejecta transport cannot be measured directly; it can only be constrained by quantifying the fresh material across Didymos' surface.

One of the goals of the HS-H instrument is to characterize the space weathering degree of each surface patch of the two asteroids. The band depth, the spectral slope, and the albedo represent the input data to test the hypothesis highlighted above.
The Q- and S-complex asteroids' spectra are characterized by the 1 $\mu$m band corresponding to olivine-pyroxene compositions. The HS-H spectral interval covers the beginning of this first band. Thus, we define the band depth by the ratio between the maximum and minimum spectral responses. 

A spectrum of the crater interior may allow direct identification of the meteoritic counterpart of Dimorphos. By finding the best meteorite analogue for the measured spectra, we can estimate the density of the body, which will provide information about the porosity properties. These findings are part of the core requirements for the ESA/Hera mission.

To accomplish the objectives mentioned above, the following data products will be generated based on HS-H acquired observations, and with the supporting data from other instruments onboard the spacecraft:
\begin{description}
    \item[Compositional maps:] these will include taxonomic classification maps, best meteorite counterpart for each surface patch, the 1 $\mu$m band depth, and band center for each sampled surface. Based on empirical models, we expect to be able to estimate the percent of mafic minerals \citep{2023A&A...669A.101K, 2024PSJ.....5...85K}. With this information, we will be able to constrain the asteroid density by considering the best meteorite analogue for the surface. In particular, the planned trajectory of ESA Hera will allow us to obtain these maps with high spatial resolution for the impact area on Dimorphos.

    \item[Space weathering maps:] the effects of space weathering will be quantified based on the spectral slope and band depth. Additionally, a principal component analysis (or similar techniques) will also be used for quantifying the weathering degree of each surface patch \citep{2018Icar..299..386K, 2007M&PS...42.1791I}. This will allow us to identify the re-deposited ejecta on Didymos, to measure its composition, and to quantify the dynamical effect it produced when it impacted on the object.

    \item[The albedo map] over the HS-H spectral interval will be computed as an average across all 25 channels.  The values will be compared with those obtained by AFC.  

    \item[The global phase curves] of the two asteroids and the opposition effect. The data obtained over the different phases of the mission will be combined to characterize the global phase curve of the asteroid. The expected phase angle variation across the imaged surfaces will be in the range from 0 to 100$^\circ$.
\end{description}

The exogenous material found on the surfaces of B- and C-type asteroids (101955) Bennu and (162173) Ryugu by the NASA/OSIRIS-REx and JAXA/Hayabusa2 space missions, respectively, triggered new hypotheses regarding the asteroids' collisional histories \citep{2021MNRAS.508.2053T}. Thus, one of the key objectives of the HS-H instrument will be to first identify and then to constrain the presence of exogenous material on the surfaces of Didymos and Dimorphos. This will be done by searching for outliers in the taxonomic classification and albedo maps mentioned above.

The main complementary instrument for HS-H is ASPECT (N\"asil\"a et al. 2024, this issue), which is onboard the Milani cubesat. This is also a hyperspectral imaging spectrometer, which has 72 channels and covers the 0.5 to 2.5~$\mu $m spectral interval. Planned common operations will improve the reliability of the obtained data products. Some limitations are imposed by the different trajectories of the spacecraft and of the cubesat. 

The images made by HS-H will be complementary to those performed by the two Asteroid Framing Cameras (AFC) developed by the company Jena Optronik. The spatial resolution at the HS-H subpixel level is comparable to that of the AFCs. These data will be used for the construction of shape models and for geomorphological studies. 


\section {Expected Investigation}
\label{section4}

The science objectives mentioned above can be achieved by taking into account the spatial resolution during each mission phase and the possible data analysis techniques. Here we provide a short summary of the different views observed by the HS-H instrument and of the toolbox currently available for the data analysis. 

\subsection{Views during various mission phases}

In order to generate the possible views of the instrument across the different phases of the mission we used the \emph{Hera SPICE Kernel Dataset}\footnote{\url{https://www.cosmos.esa.int/web/spice/spice-for-hera}} \citep{esa_spice_operational_skd_2025} and the \emph{shapeViewer} scientific software\footnote{\url{https://www.comet-toolbox.com/shapeViewer.html}} \citep{2012P&SS...66...79V}. 

\begin{figure*}
    \centering
    \includegraphics[width=0.95\linewidth]{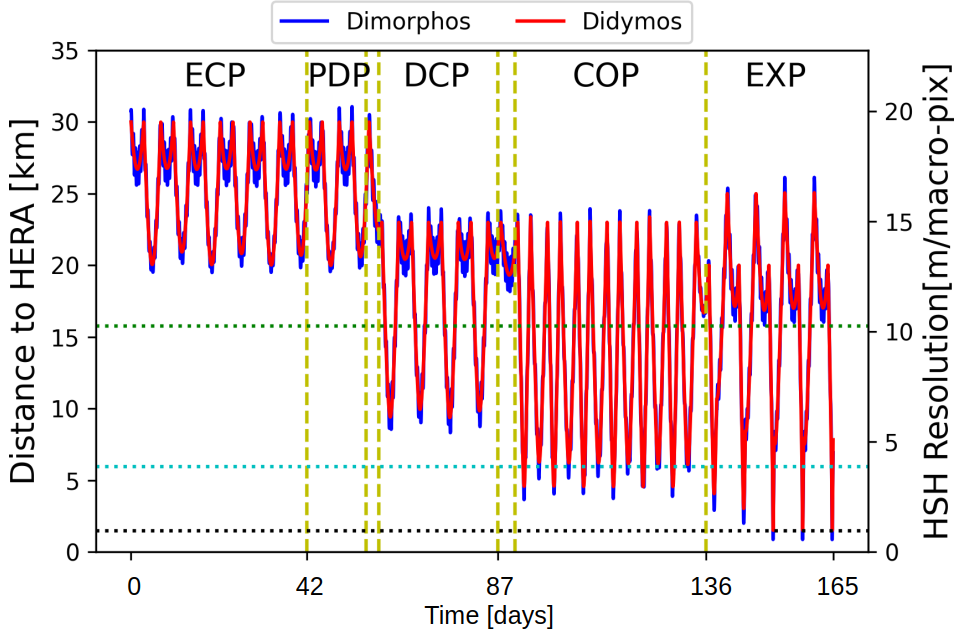}
    \caption{ The HS-H spatial resolution during asteroid observations. The left Oy axis shows the distance between the Hera spacecraft and Didymos (red curve) and Dimorphos(blue curve). The right Oy axis shows the spatial resolution for imaging the surface of two asteroids. The values are given as m/macropixel. The vertical yellow markers delimit the different mission phases. The horizontal black, cyan, and green markers show when Dimorphos (black), Dydimos (cyan), and the entire system (green) fit in the small axis (217 macropixels) of the instrument. The acronyms are explained in the text.}
    \label{MissionPhases}
\end{figure*}

We also used the SPICE toolkit\footnote{\url{https://naif.jpl.nasa.gov/naif/toolkit.html}} \citep{1996P&SS...44...65A, 2018P&SS..150....9A} to plot the observing geometry (distance, phase angle) over the different mission phases.   The spatial resolution of the HS-H over the mission asteroid phases in shown in  Fig.~\ref{MissionPhases}. 

\begin{figure}
    \centering
    \includegraphics[width=0.95\linewidth]{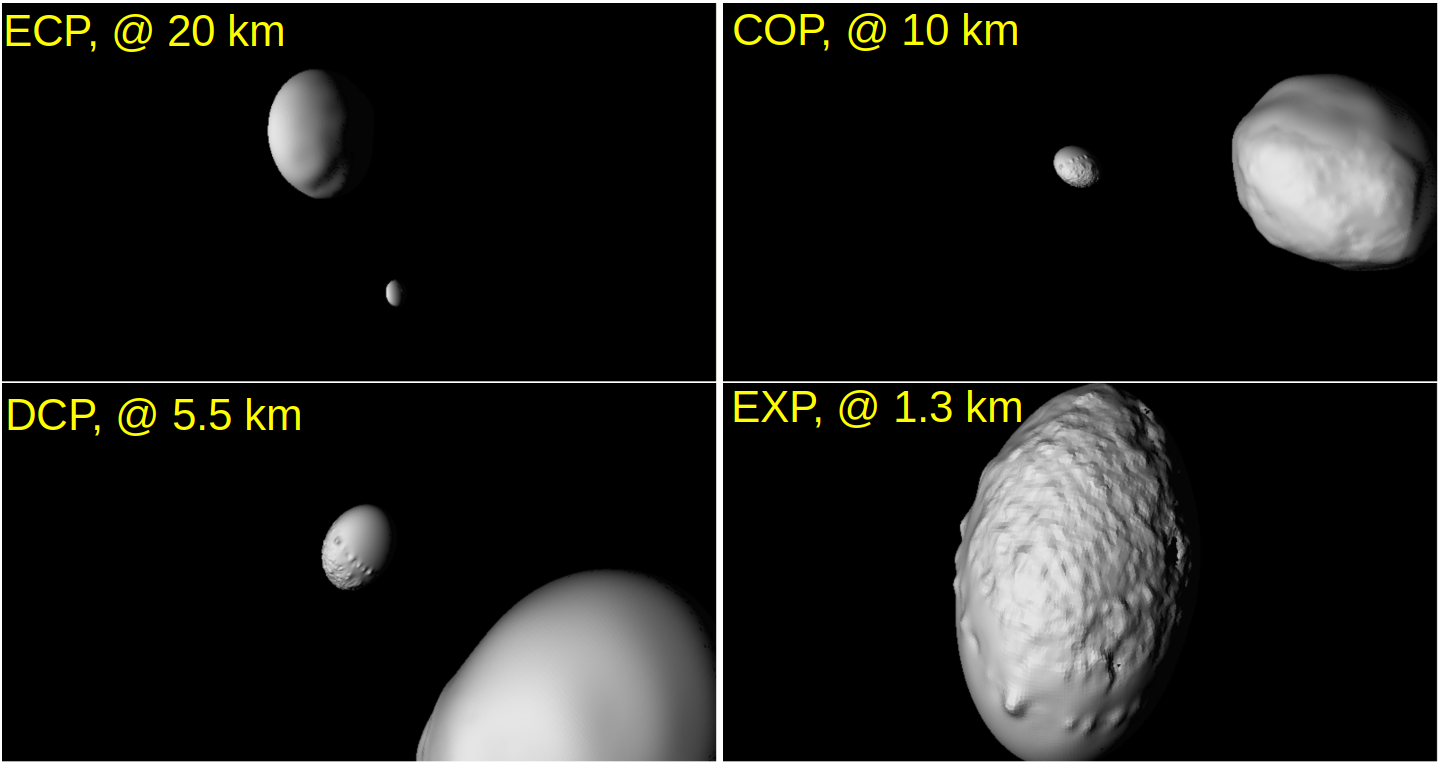}
    \caption{ Examples of HS-H field of view during different phases of the mission. The distance between the Hera spacecraft and Dimorphos center is provided. The closest approaches during each mission phase were considered.  The images were obtained using \emph{shapeViewer} software provided by Jean-Baptiste Vincent. 
    }
    \label{fig:HSHviews}
\end{figure}

The Hera proximity operations on the asteroid consist of several mission phases. Some examples of how Didymos and Dimorphos will appear in the field of view of HS-H are shown in  Fig.~\ref{fig:HSHviews}. This figure is an example of the possible schedule of the spacecraft operations at the asteroid. The exact time periods of each phase may be slightly updated before the arrival.

Following a safe rendezvous sequence, Hera’s arrival at (65803) Didymos is expected in the autumn of 2026. This marks the beginning of the Early Characterization Phase (ECP) of Proximity Operations. During ECP, the spacecraft will be at a distance between about 40 and 20~km from Dimorphos. The spatial resolution of HS-H will be around 15 m/macropixel and the entire binary system Didymos-Dimorphos will be in the field of view of the instrument. For HS-H, this will be the best time to characterize the dynamics of the system, to scrutinize the dust environment, and to search for the presence of boulders \citep{2025PSJ.....6..155F}. It will also allow to obtain global average spectra of both asteroids to compare them with the ground-based observations \citep{2023PSJ.....4..229P, 2022PSJ.....3..183I, 2023LPICo2806.2080L} or with the data obtained from the LUKE instrument on the DART-LICIACube cubesat \citep{2024Natur.627..505D}.

After the ECP, the Payload Deployment (PDP) will take place. Both cubesats, Juventas and Milani, will be released, and their instruments will be commissioned. During this mission phase, HS-H will perform simultaneous observations with ASPECT in order to cross-calibrate both instruments. 

A transition phase will follow before the Detailed Characterization Phase (DCP).  During the DCP time frame, Hera will be at a minimum distance of 10~km from Dimorphos, and during the four planned close approaches, the resolution of HS-H will be close to 5~m/macropixel. The observations phase angle will vary between about $70^\circ$ and $0^\circ$, thus we will be able to characterize the opposition effect.   

The next phase of the mission is the Close Operations Phase (COP), when Hera will make close approaches to Dimorphos down to 4.5 km. At this distance, HS-H  spatial resolution will be about 2.5~m/macropixels (thus, 0.5~m/subpixel). Simultaneous observations between the HS-H and other instruments are planned.  During this phase, the cubesats will land, an event planned to also be imaged by HS-H. During COP, we will also have the opportunity to obtain data at large phase angles.

The Experimental (EXP) phase is the last phase of the mission. During the planned close approaches (five very close fly-bys are planned), the HS-H will be able to take images with high spatial resolution of the DART impact place, less than about 1~m/macropixel (thus, 0.2~m/subpixel). 

\subsection{Data analysis toolbox}

Following the completion of all calibration procedures, data analysis proceeds in two steps. The first is demosaicing, which disentangles the spatial and spectral information of each image. For the subsequent step, we provide some of the spectral analysis techniques that can be applied to the HS-H data.

\subsubsection{Demosaicing}

Because of the arrangement of $5\times5$ subpixels into one macropixel, see section~\ref{technicalspecification}, at each subpixel only information for one of the 25 wavelength bins is available. Earlier versions of the HyperScout instrument were targeted at Earth observation and were supposed to be operated in a pushbroom mode that accumulates information over different wavelength bins for a certain patch on the Earth's surface. While this may to some extent also be possible for the Hera mission, the irregular shape of the target bodies makes it more complicated to co-register different images and we should be prepared to extract as much information as possible from individual HS-H images. The process of replenishing the full data cube of a hyperspectral imager from an image providing only the brightness for one wavelength at each pixel is called demosaicing.

The simplest demosaicing approach is just assembling the spectral values of all subpixels for each macropixel. This does, of course, reduce the spatial resolution to the macropixels and moreover, since each subpixel corresponds to a different surface patch, the obtained spectrum could only be correct if it was homogeneous over the macropixel. Therefore, spectra obtained in this way are, in general, relatively noisy. However, this simple procedure is the least affected by artifacts introduced by more complicated numerical processing methods and is appropriate to obtain the mean spectrum for a larger area comprising several macropixels.

More advanced traditional methods are based on interpolation, such as Bayer interpolation \citep{mihoubi2018snapshot}, which de facto also reduces the resolution. The most recent techniques are based on neural network algorithms \citep{dijkstra2018hyperspectral}, but these require a large dataset for training.  

We have developed a family of novel demosaicing algorithms for the HS-H images in order to replenish the full HS-H data cube with two spatial dimensions at subpixel resolution and one spectral dimension.
%
%
If we have only one individual HS-H image and no additional information at all, the problem of reconstructing the full data cube is completely unconstrained. To illustrate this, we can write the brightness $f$ at the subpixel in row $i$ and column $j$ in wavelength bin $k$ as the product of a brightness scaling factor $b$ and a normalized spectrum $\hat{f}$:
\begin{equation}
f(i,j,k)=b(i,j)\,\hat{f}(i,j,k)
\label{DM1}
\end{equation}
with
\begin{equation}
\frac{1}{25} \sum_{k=1}^{25} \hat{f}(i,j,k) = 1 .
\label{DM2}
\end{equation}
While sometimes spectra are normalized to the same value at some particular wavelength for comparative plots, we here normalize the spectra to the same mean value in order to obtain unbiased brightness scaling factors.

At any subpixel $(i,j)$, we can assume an arbitrary normalized spectrum $\hat{f}$ and reconcile it with the measured value $f$ (only available for one wavelength bin $k$) by choosing an appropriate brightness factor $b$. Therefore, additional assumptions are needed to constrain the problem. If we just collect the values of all subpixels over one macropixel (or interpolate between respective subpixels in different macropixels), this implies that we assume that normalized spectra are spatially smooth on the macropixel scale. This may indeed be the case for many regions of the asteroid. However, this simple approach also implies the assumption that the brightness factor is spatially smooth on the macropixel scale, and this may generally not be the case because of local albedo variations, and moreover because of varying shading due to surface roughness. This yields noisy spectra and requires averaging over many macropixels, implying the assumption that normalized spectra are spatially smooth even on the scale of many macropixels.

Our novel demosaicing approaches require fewer far-reaching constraints. One core idea is to use measured brightness \emph{ratios} of two different wavelength bins at spatially adjacent pixels. If the brightness factors of adjacent pixels are similar (i.\,e., the brightness factors are spatially smooth on \emph{subpixel} scale), the measured ratios provide an estimate of the respective ratios in the normalized
spectrum. Our different demosaicing approaches will be described in full detail in a subsequent article.

\subsubsection{Taxonomic classification}

Once the spectra are obtained, the next approach is to characterize the surface covered by each pixel. To compare the HS-H spectrum with other spectra, we must consider the narrow-band transmission functions of HS-H. The fact that their shapes are not uniform in the in-band and they have out-of-band ripples requires applying these functions to any comparison spectrum (Eq.~\ref{eqhalogenlampsspectrum}). Then, for each input data, we get the equivalent HS-H observed spectrum containing 25 reflectance values computed for each one of the HS-H bands.

The taxonomic classification task can be achieved using machine learning techniques \citep{2023A&A...669A.101K}. The input data is represented by the 25 reflectance values, and the output is the taxonomic type. Thus, the learning problem is based on finding the best multi-class classifier. This is typically achieved by training a classifier to predict the probability distribution over the classes by minimizing the cross-entropy loss, which is a measure of the dissimilarity between the predicted probability distribution and the true distribution.

The dataset described by \citep{2022A&A...665A..26M} was used for training our algorithms. It contains up to 2983 spectral samples of 2125 asteroids. We select 11 classes that are the most relevant for our case study. We split the dataset such that 60$\%$ of the data is used for training, 20$\%$ for validation, and 20$\%$ for testing. A stratified split was performed, meaning that there are enough samples of each taxonomic class in each set. Each spectrum is interpolated by a spline curve to obtain a fit for the reflectance spectrum and to apply the transmission functions of HS-H. The data is augmented as needed by small shifts or rotations, or by adding noise similar to the one introduced HS-H instrument.

We implement the traditional machine learning algorithms \citep{2020SciPy-NMeth} such as k-Nearest Neighbors, Decision Tree, and Random Forest, which we optimize via grid search. Also, we try different dense (DNN) and convolutional (CNN) neural network architectures, aiming to get the best accuracy by width and depth scaling. A comprehensive analysis of these methods is done to choose the best classifier and obtain accurate classification maps.


\section{Proof of concept: the HS-H laboratory spectra of meteorites}
\label{section5}

Two bulk meteorite samples were received from Archaeology, Environmental Changes, and Geo-Chemistry (AMGC) in Belgium. One piece, approximately $6\times3$~cm in size, is a fragment from El Hammami meteorite, an H5 ordinary chondrite. The other sample, measuring about $4\times3$~cm, was part of the Sayh al Uhaymir (SaU) 001 meteorite, also an ordinary chondrite but of L5 subtype. The two samples are compared in Table~\ref{tbl:meteorites}. Both pieces are polished.

\begin{table}[]
\caption{The description of the meteorite samples used for validating the HS-H instrument and the methods.}
\begin{tabular}{|c|c|c|} \hline
                                              & \textbf{Sample \#1}         & \textbf{Sample \#2}         \\ \hline
\multicolumn{1}{|c|}{\textbf{Meteorite Name}} & El Hammami                  & SaU 001                     \\ \hline
\multicolumn{1}{|c|}{\textbf{Meteorite Type}} & H5                          & L5                          \\ \hline
\multicolumn{1}{|c|}{\textbf{Weight (g)}}     & 21.70                       & 15.02                       \\ \hline
\multicolumn{1}{|c|}{\textbf{Size (cm $\times$ cm)}} & $\sim \ 6 \times 3$ & $\sim \ 4 \times 3$ \\ \hline
\end{tabular}
\label{tbl:meteorites}
\end{table}

\begin{figure}
    \centering
    \includegraphics[width=0.95\linewidth]{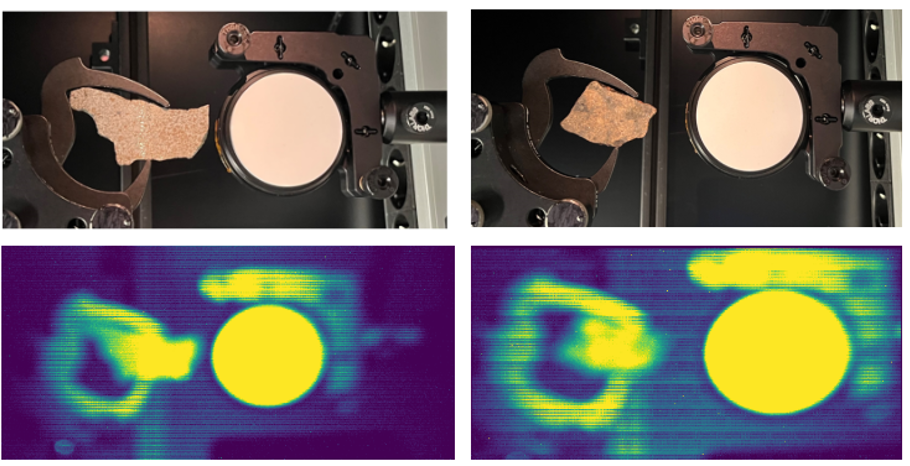}
    \caption{Comparison between the images obtained with a camera for photography and the raw images obtained with HS-H instrument for the two samples. The appearance of horizontal and vertical stripes in the HS-H images is due to the differing sensitivities of the subpixels.
}
    \label{fig:meteorites}
\end{figure}

Each meteorite sample was positioned above a 2~inch diameter Spectralon 99\% diffuse target with a calibrated and certified reflectance standard (SRS-99-020). A halogen lamp, Mikropack HL-2000-FHSA-HP (Serial No. 034990084), was placed at a distance of 40~cm from the targets at a $30^\circ$ angle  to ensure a uniform illumination of both the meteorite and the Spectralon surface. The targets were positioned 103~cm from the HS-H $+z$~axis to slightly defocus the target. All measurements were conducted with the lab room lights turned off.

A total of 45 images were acquired with HS-H, 15 for El Hammami and 30 for SaU 001.  To simplify the acquisition process, data was recorded only for a region of interest (ROI) of $512 \times 1056$ pixels, which helped in reducing the necessary read-out times. The experimental set-up (meteorite sample and lamp) can be seen in Fig.~\ref{fig:meteorites}. Images acquired by HS-H are provided for comparison.

The exposure times of both samples were 500~ms, 1~sec, and 2~sec.  However, the results presented here were obtained using a single image for each of the two meteorites with an exposure time of 1 sec.  The signal on the meteorite's surfaces is on the order of 1000~DN, which corresponds to an SNR of about~60. Although the images obtained with 2~sec exposure have a better SNR for the meteorite spectra, the subpixels comprising the Spectralon were saturated.

Post-calibration of the meteorite spectral targets was conducted in the laboratory with a dedicated spectrometer. The reference spectra were obtained over the wavelength range 0.550--4.200~$\mu$m with a dedicated setup using the SHADOWS spectro-goniometer at Institut de Plan\'{e}tologie et d'Astrophysique de Grenoble (IPAG), France. The sample area measured was about $5.2 \times 5.2$~mm${}^2$ covered with 7 optical fibers with a cross section of about $1.3 \times 1.7$~mm${}^2$ each. The optical arm of the spectro-goniometer was set at the nominal geometry of incidence $0^\circ$ and emergence $30^\circ$. The HS-H transfer functions, Eq.~(\ref{eqhalogenlampsspectrum}), were applied to these reference spectra to perform the comparison.

The data reduction for the HS-H measurements first included the dark and flat field corrections. Then, we retrieved the Spectralon spectrum in order to quantify the spectral curve of the source of light. Finally, we select various regions across the meteorite surface.  To determine the meteorite reflectance spectrum, we divided the raw HS-H meteorite spectrum by the spectralon one.
\begin{figure*}
    \centering
    \includegraphics[width=0.45\linewidth]{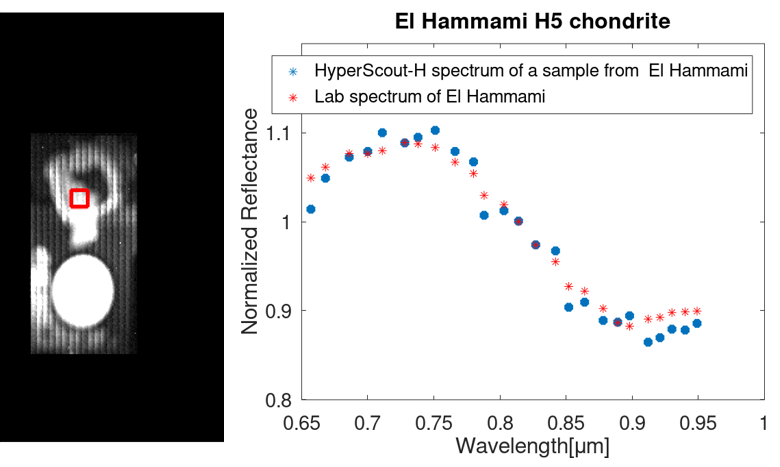}
    \includegraphics[width=0.45\linewidth]{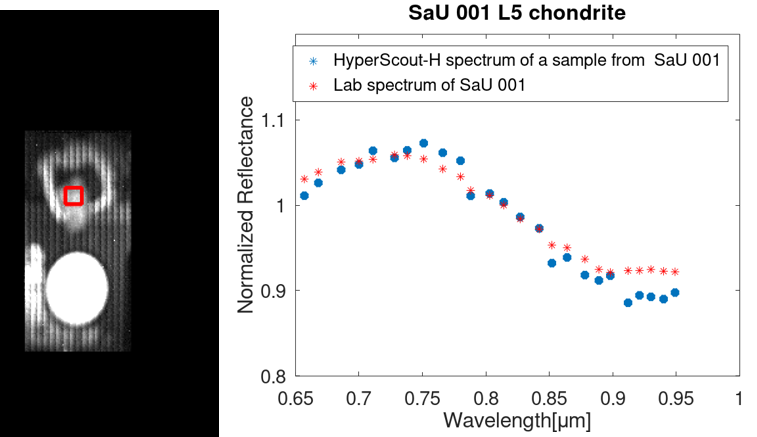}
    \caption{The comparison between the average meteorite spectra obtained with HS-H over a region of 30 $\times$ 30 macropixels (outlined on the left) and the reference spectra obtained with SHADOWS spectro-goniometer for the same samples.}
    \label{fig:meteorites_cross}
\end{figure*}
In Fig.~\ref{fig:meteorites_cross}, the average meteorite spectra obtained with HS-H over a region of 30 $\times$ 30 $macropixels$ and the reference spectra obtained with SHADOWS spectro-goniometer are shown. This serves as a laboratory proof of concept for the instrument.

\begin{figure*}
    \centering
    \includegraphics[width=0.95\linewidth]{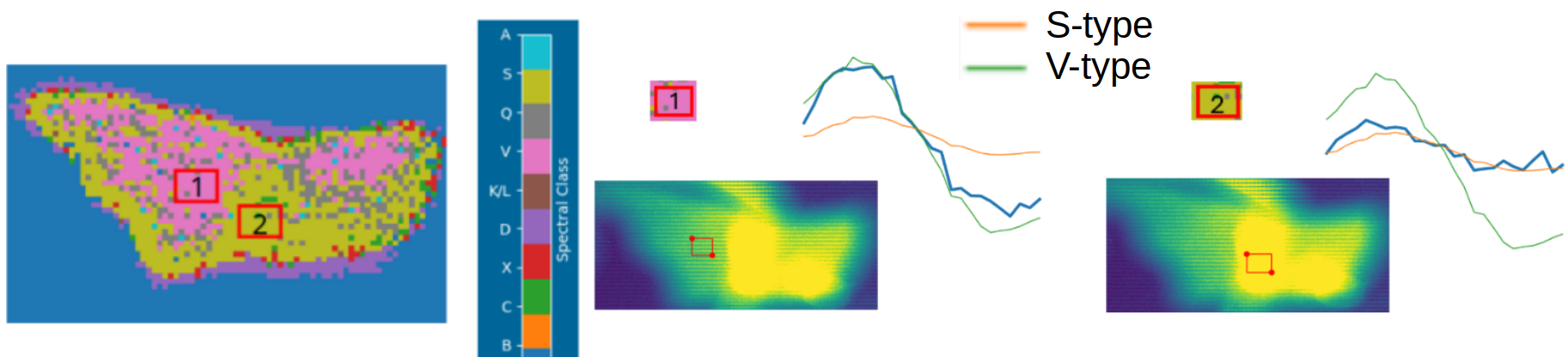}
    \caption{ Spectral variation across the meteorite surface. The spectrum corresponding to each macropixel was classified using the approach described in Section 4. Two different regions are highlighted for comparison.}
    \label{fig:meteorites_variation}
\end{figure*}

The spatial resolution of the HS-H measurements was 0.68~mm/macropixel.  Spectral variation was determined across the meteorite surface by using taxonomic classification. Consistent patterns are highlighted in Fig.~\ref{fig:meteorites_variation}, which validates our data analysis approach. The S-type and the V-type represent the template spectra for these classes as defined by \citet{2022A&A...665A..26M}. This classification highlights the spectral differences in band depth and band minima within the 1~$\mu$m band, which is characteristic of olivine-pyroxene compositions.

\section{Conclusions}
\label{section6}
The HS-H instrument is set to fly aboard the ESA/Hera spacecraft. Its observations will provide key insights into the Didymos-Dimorphos composition, space-weathering effects, and the possible presence of exogenous material. Additionally, the HS-H was used during the spacecraft's cruise phase in March 2025, when Hera performed a flyby of Mars. During this phase, the Martian moon Deimos could be spectrally characterized by the HS-H.

The HS-H captures hyperspectral images within the 0.65--0.95~$\mu$m wavelength range. Each macropixel consists of a $5 \times 5$ matrix of pixels and records signals across 25 distinct narrow bands, with each band corresponding to an individual pixel. Pre-flight calibration of the instrument has already been completed, and its capabilities were demonstrated using two meteorite samples as proof of concept.

The data gathered from the HS-H allows the creation of 2D maps that highlight various spectral features, such as taxonomic classification, spectral slope, and band parameters. To analyze these observations efficiently, a dedicated software toolbox was developed specifically for processing the instrument’s data.

\backmatter



\section*{Declarations}


\begin{itemize}
\item Funding:
The authors MP, JdL, GPP, GK, BVN, PM received partial funding from ESA Contract No. 4000144275/24/NL/GLC/my. LKR thanks the FWO for the funding Postdoctoral Fellowship Junior 1295024N.
\item Conflict of interest/Competing interests:
The authors, M.E. and N.V., are employed by \emph{cosine Remote Sensing BV}.

\end{itemize}







\begin{appendices}






\end{appendices}


\bibliography{hsh}

\begin{thebibliography}{70}
\providecommand{\natexlab}[1]{#1}
\providecommand{\url}[1]{{#1}}
\providecommand{\urlprefix}{URL }
\providecommand{\doi}[1]{\url{https://doi.org/#1}}
\providecommand{\eprint}[2][]{\url{#2}}
 \bibcommenthead

\bibitem[{{Abell} et~al(2015){Abell}, {Barbee}, {Chodas}, {Kawaguchi}, {Landis}, {Mazanek}, and {Michel}}]{2015aste.book..855A}
{Abell} PA, {Barbee} BW, {Chodas} PW, et~al (2015) {Human Exploration of Near-Earth Asteroids}. In: {Michel} P, {DeMeo} FE, {Bottke} WF (eds) Asteroids IV. p 855--880, \doi{10.2458/azu_uapress_9780816532131-ch043}

\bibitem[{{Acton} et~al(2018){Acton}, {Bachman}, {Semenov}, and {Wright}}]{2018P&SS..150....9A}
{Acton} C, {Bachman} N, {Semenov} B, et~al (2018) {A look towards the future in the handling of space science mission geometry}. Planetary and Space Science 150:9--12. \doi{10.1016/j.pss.2017.02.013}

\bibitem[{{Acton}(1996)}]{1996P&SS...44...65A}
{Acton} CH (1996) {Ancillary data services of NASA's Navigation and Ancillary Information Facility}. Planetary and Space Science 44(1):65--70. \doi{10.1016/0032-0633(95)00107-7}

\bibitem[{{Alvarez} et~al(1980){Alvarez}, {Alvarez}, {Asaro}, and {Michel}}]{1980Sci...208.1095A}
{Alvarez} LW, {Alvarez} W, {Asaro} F, et~al (1980) {Extraterrestrial Cause for the Cretaceous-Tertiary Extinction}. Science 208(4448):1095--1108. \doi{10.1126/science.208.4448.1095}

\bibitem[{{Barnouin} et~al(2024){Barnouin}, {Ballouz}, {Marchi}, {Vincent}, {Agrusa}, {Zhang}, {Ernst}, {Pajola}, {Tusberti}, {Lucchetti}, {Daly}, {Palmer}, {Walsh}, {Michel}, {Sunshine}, {Rizos}, {Farnham}, {Richardson}, {Parro}, {Murdoch}, {Robin}, {Hirabayashi}, {Kahout}, {Asphaug}, {Raducan}, {Jutzi}, {Ferrari}, {Hasselmann}, {CampoBagatin}, {Chabot}, {Li}, {Cheng}, {Nolan}, {Stickle}, {Karatekin}, {Dotto}, {Della Corte}, {Mazzotta Epifani}, {Rossi}, {Gai}, {Deshapriya}, {Bertini}, {Zinzi}, {Trigo-Rodriguez}, {Beccarelli}, {Ivanovski}, {Brucato}, {Poggiali}, {Zanotti}, {Amoroso}, {Capannolo}, {Cremonese}, {Dall'Ora}, {Ieva}, {Impresario}, {Lavagn}, {Modenini}, {Palumbo}, {Perna}, {Pirrotta}, {Tortora}, {Zannoni}, and {Rivkin}}]{2024NatCo..15.6202B}
{Barnouin} O, {Ballouz} RL, {Marchi} S, et~al (2024) {The geology and evolution of the Near-Earth binary asteroid system (65803) Didymos}. Nature Communications 15:6202. \doi{10.1038/s41467-024-50146-x}

\bibitem[{{Benhadj} et~al(2024){Benhadj}, {Livens}, {Esposito}, {Vercruyssen}, {Van Dijk}, {Soukup}, {Zuccaro}, and {Maresi}}]{2024IJRS...45.2488B}
{Benhadj} I, {Livens} S, {Esposito} M, et~al (2024) {HyperScout-1 inflight calibration and product validation}. International Journal of Remote Sensing 45(7):2488--2513. \doi{10.1080/01431161.2024.2331979}

\bibitem[{{Binzel} et~al(2010){Binzel}, {Morbidelli}, {Merouane}, {DeMeo}, {Birlan}, {Vernazza}, {Thomas}, {Rivkin}, {Bus}, and {Tokunaga}}]{2010Natur.463..331B}
{Binzel} RP, {Morbidelli} A, {Merouane} S, et~al (2010) {Earth encounters as the origin of fresh surfaces on near-Earth asteroids}. \nat 463(7279):331--334. \doi{10.1038/nature08709}

\bibitem[{{Binzel} et~al(2019){Binzel}, {DeMeo}, {Turtelboom}, {Bus}, {Tokunaga}, {Burbine}, {Lantz}, {Polishook}, {Carry}, {Morbidelli}, {Birlan}, {Vernazza}, {Burt}, {Moskovitz}, {Slivan}, {Thomas}, {Rivkin}, {Hicks}, {Dunn}, {Reddy}, {Sanchez}, {Granvik}, and {Kohout}}]{2019Icar..324...41B}
{Binzel} RP, {DeMeo} FE, {Turtelboom} EV, et~al (2019) {Compositional distributions and evolutionary processes for the near-Earth object population: Results from the MIT-Hawaii Near-Earth Object Spectroscopic Survey (MITHNEOS)}. \icarus 324:41--76. \doi{10.1016/j.icarus.2018.12.035}, {\href{https://arxiv.org/abs/2004.05090}{{arXiv:2004.05090}}} {[astro-ph.EP]}

\bibitem[{Borengasser et~al(2008)Borengasser, Hungate, and Watkins}]{Borengasser}
Borengasser M, Hungate WS, Watkins R (2008) Hyperspectral Remote Sensing Principles and Applications. CRC Press Taylor and Francis Group, Indiana State University, Terre Haute, Indiana, U.S.A

\bibitem[{{Borovi{\v{c}}ka} et~al(2013){Borovi{\v{c}}ka}, {Spurn{\'y}}, {Brown}, {Wiegert}, {Kalenda}, {Clark}, and {Shrben{\'y}}}]{2013Natur.503..235B}
{Borovi{\v{c}}ka} J, {Spurn{\'y}} P, {Brown} P, et~al (2013) {The trajectory, structure and origin of the Chelyabinsk asteroidal impactor}. \nat 503(7475):235--237. \doi{10.1038/nature12671}

\bibitem[{Bottke et~al(2002)Bottke, Morbidelli, Jedicke, Petit, Levison, Michel, and Metcalfe}]{2002Icar..156..399B}
Bottke WF, Morbidelli A, Jedicke R, et~al (2002) Debiased {O}rbital and {A}bsolute {M}agnitude {D}istribution of the {N}ear-{E}arth {O}bjects. Icarus 156:399--433. \doi{10.1006/icar.2001.6788}

\bibitem[{{Brown} et~al(2013){Brown}, {Assink}, {Astiz}, {Blaauw}, {Boslough}, {Borovi{\v{c}}ka}, {Brachet}, {Brown}, {Campbell-Brown}, {Ceranna}, {Cooke}, {de Groot-Hedlin}, {Drob}, {Edwards}, {Evers}, {Garces}, {Gill}, {Hedlin}, {Kingery}, {Laske}, {Le Pichon}, {Mialle}, {Moser}, {Saffer}, {Silber}, {Smets}, {Spalding}, {Spurn{\'y}}, {Tagliaferri}, {Uren}, {Weryk}, {Whitaker}, and {Krzeminski}}]{2013Natur.503..238B}
{Brown} PG, {Assink} JD, {Astiz} L, et~al (2013) {A 500-kiloton airburst over Chelyabinsk and an enhanced hazard from small impactors}. \nat 503(7475):238--241. \doi{10.1038/nature12741}

\bibitem[{{Brunetto} et~al(2015){Brunetto}, {Loeffler}, {Nesvorn{\'y}}, {Sasaki}, and {Strazzulla}}]{2015aste.book..597B}
{Brunetto} R, {Loeffler} MJ, {Nesvorn{\'y}} D, et~al (2015) {Asteroid Surface Alteration by Space Weathering Processes}. In: {Michel} P, {DeMeo} FE, {Bottke} WF (eds) Asteroids IV. p 597--616, \doi{10.2458/azu_uapress_9780816532131-ch031}

\bibitem[{{Chabot} et~al(2024){Chabot}, {Rivkin}, {Cheng}, {Barnouin}, {Fahnestock}, {Richardson}, {Stickle}, {Thomas}, {Ernst}, {Terik Daly}, {Dotto}, {Zinzi}, {Chesley}, {Moskovitz}, {Barbee}, {Abell}, {Agrusa}, {Bannister}, {Beccarelli}, {Bekker}, {Bruck Syal}, {Buratti}, {Busch}, {Campo Bagatin}, {Chatelain}, {Chocron}, {Collins}, {Conversi}, {Davison}, {DeCoster}, {Prasanna Deshapriya}, {Eggl}, {Espiritu}, {Farnham}, {Ferrais}, {Ferrari}, {F{\"o}hring}, {Fuentes-Mu{\~n}oz}, {Gai}, {Giordano}, {Glenar}, {Gomez}, {Graninger}, {Green}, {Greenstreet}, {Hasselmann}, {Herreros}, {Hirabayashi}, {Hus{\'a}rik}, {Ieva}, {Ivanovski}, {Jackson}, {Jehin}, {Jutzi}, {Karatekin}, {Knight}, {Kolokolova}, {Kumamoto}, {K{\"u}ppers}, {La Forgia}, {Lazzarin}, {Li}, {Lister}, {Lolachi}, {Lucas}, {Lucchetti}, {Luther}, {Makadia}, {Mazzotta Epifani}, {McMahon}, {Merisio}, {Merrill}, {Meyer}, {Michel}, {Micheli}, {Migliorini}, {Minker}, {Modenini}, {Moreno}, {Murdoch}, {Murphy}, {Naidu}, {Nair}, {Nakano}, {Opitom}, {Orm{\"o}},
  {Michael Owen}, {Pajola}, {Palmer}, {Palumbo}, {Panicucci}, {Parro}, {Pearl}, {Penttil{\"a}}, {Perna}, {Petrescu}, {Pravec}, {Raducan}, {Ramesh}, {Ridden-Harper}, {Rizos}, {Rossi}, {Roth}, {Ro{\.z}ek}, {Rozitis}, {Ryan}, {Ryan}, {S{\'a}nchez}, {Santana-Ros}, {Scheeres}, {Scheirich}, {Senel}, {Snodgrass}, {Soldini}, {Souami}, {Statler}, {Street}, {Stubbs}, {Sunshine}, {Tan}, {Tancredi}, {Tinsman}, {Tortora}, {Tusberti}, {Walker}, {Waller}, {W{\"u}nnemann}, {Zannoni}, and {Zhang}}]{2024PSJ.....5...49C}
{Chabot} NL, {Rivkin} AS, {Cheng} AF, et~al (2024) {Achievement of the Planetary Defense Investigations of the Double Asteroid Redirection Test (DART) Mission}. \psj 5(2):49. \doi{10.3847/PSJ/ad16e6}

\bibitem[{{Cheng} et~al(2015){Cheng}, {Atchison}, {Kantsiper}, {Rivkin}, {Stickle}, {Reed}, {Galvez}, {Carnelli}, and {Michel}}]{2015AcAau.115..262C}
{Cheng} AF, {Atchison} JA, {Kantsiper} B, et~al (2015) {Asteroid impact and deflection assessment mission}. Acta Astronautica 115:262--269. \doi{10.1016/j.actaastro.2015.05.021}

\bibitem[{{Cheng} et~al(2023){Cheng}, {Agrusa}, {Barbee}, {Meyer}, {Farnham}, {Raducan}, {Richardson}, {Dotto}, {Zinzi}, {Della Corte}, {Statler}, {Chesley}, {Naidu}, {Hirabayashi}, {Li}, {Eggl}, {Barnouin}, {Chabot}, {Chocron}, {Collins}, {Daly}, {Davison}, {DeCoster}, {Ernst}, {Ferrari}, {Graninger}, {Jacobson}, {Jutzi}, {Kumamoto}, {Luther}, {Lyzhoft}, {Michel}, {Murdoch}, {Nakano}, {Palmer}, {Rivkin}, {Scheeres}, {Stickle}, {Sunshine}, {Trigo-Rodriguez}, {Vincent}, {Walker}, {W{\"u}nnemann}, {Zhang}, {Amoroso}, {Bertini}, {Brucato}, {Capannolo}, {Cremonese}, {Dall'Ora}, {Deshapriya}, {Gai}, {Hasselmann}, {Ieva}, {Impresario}, {Ivanovski}, {Lavagna}, {Lucchetti}, {Epifani}, {Modenini}, {Pajola}, {Palumbo}, {Perna}, {Pirrotta}, {Poggiali}, {Rossi}, {Tortora}, {Zannoni}, and {Zanotti}}]{2023Natur.616..457C}
{Cheng} AF, {Agrusa} HF, {Barbee} BW, et~al (2023) {Momentum transfer from the DART mission kinetic impact on asteroid Dimorphos}. \nat 616(7957):457--460. \doi{10.1038/s41586-023-05878-z}, {\href{https://arxiv.org/abs/2303.03464}{{arXiv:2303.03464}}} {[astro-ph.EP]}

\bibitem[{{Chyba} et~al(1993){Chyba}, {Thomas}, and {Zahnle}}]{1993Natur.361...40C}
{Chyba} CF, {Thomas} PJ, {Zahnle} KJ (1993) {The 1908 Tunguska explosion: atmospheric disruption of a stony asteroid}. \nat 361(6407):40--44. \doi{10.1038/361040a0}

\bibitem[{{Clark} et~al(2002){Clark}, {Hapke}, {Pieters}, and {Britt}}]{2002aste.book..585C}
{Clark} BE, {Hapke} B, {Pieters} C, et~al (2002) {Asteroid Space Weathering and Regolith Evolution}. In: {Bottke} JW.~F., {Cellino} A, {Paolicchi} P, et~al (eds) Asteroids III. p 585--599

\bibitem[{{Daly} et~al(2023){Daly}, {Ernst}, {Barnouin}, {Chabot}, {Rivkin}, {Cheng}, {Adams}, {Agrusa}, {Abel}, {Alford}, {Asphaug}, {Atchison}, {Badger}, {Baki}, {Ballouz}, {Bekker}, {Bellerose}, {Bhaskaran}, {Buratti}, {Cambioni}, {Chen}, {Chesley}, {Chiu}, {Collins}, {Cox}, {DeCoster}, {Ericksen}, {Espiritu}, {Faber}, {Farnham}, {Ferrari}, {Fletcher}, {Gaskell}, {Graninger}, {Haque}, {Harrington-Duff}, {Hefter}, {Herreros}, {Hirabayashi}, {Huang}, {Hsieh}, {Jacobson}, {Jenkins}, {Jensenius}, {John}, {Jutzi}, {Kohout}, {Krueger}, {Laipert}, {Lopez}, {Luther}, {Lucchetti}, {Mages}, {Marchi}, {Martin}, {McQuaide}, {Michel}, {Moskovitz}, {Murphy}, {Murdoch}, {Naidu}, {Nair}, {Nolan}, {Orm{\"o}}, {Pajola}, {Palmer}, {Peachey}, {Pravec}, {Raducan}, {Ramesh}, {Ramirez}, {Reynolds}, {Richman}, {Robin}, {Rodriguez}, {Roufberg}, {Rush}, {Sawyer}, {Scheeres}, {Scheirich}, {Schwartz}, {Shannon}, {Shapiro}, {Shearer}, {Smith}, {Steele}, {Steckloff}, {Stickle}, {Sunshine}, {Superfin}, {Tarzi}, {Thomas}, {Thomas},
  {Trigo-Rodr{\'\i}guez}, {Tropf}, {Vaughan}, {Velez}, {Waller}, {Wilson}, {Wortman}, and {Zhang}}]{2023Natur.616..443D}
{Daly} RT, {Ernst} CM, {Barnouin} OS, et~al (2023) {Successful kinetic impact into an asteroid for planetary defence}. \nat 616(7957):443--447. \doi{10.1038/s41586-023-05810-5}, {\href{https://arxiv.org/abs/2303.02248}{{arXiv:2303.02248}}} {[astro-ph.EP]}

\bibitem[{{Daly} et~al(2024){Daly}, {Ernst}, {Barnouin}, {Gaskell}, {Nair}, {Agrusa}, {Chabot}, {Cheng}, {Dotto}, {Mazzotta Epifani}, {Espiritu}, {Farnham}, {Palmer}, {Pravec}, {Rivkin}, {Waller}, {Zinzi}, {DART Team}, and {LICIACube Team}}]{2024PSJ.....5...24D}
{Daly} RT, {Ernst} CM, {Barnouin} OS, et~al (2024) {An Updated Shape Model of Dimorphos from DART Data}. \psj 5(1):24. \doi{10.3847/PSJ/ad0b07}

\bibitem[{{de Le{\'o}n} et~al(2006){de Le{\'o}n}, {Licandro}, {Duffard}, and {Serra-Ricart}}]{2006AdSpR..37..178D}
{de Le{\'o}n} J, {Licandro} J, {Duffard} R, et~al (2006) {Spectral analysis and mineralogical characterization of 11 olivine pyroxene rich NEAs}. Advances in Space Research 37(1):178--183. \doi{10.1016/j.asr.2005.05.074}

\bibitem[{{de Le{\'o}n} et~al(2010){de Le{\'o}n}, {Licandro}, {Serra-Ricart}, {Pinilla-Alonso}, and {Campins}}]{2010A&A...517A..23D}
{de Le{\'o}n} J, {Licandro} J, {Serra-Ricart} M, et~al (2010) {Observations, compositional, and physical characterization of near-Earth and Mars-crosser asteroids from a spectroscopic survey}. \aap 517:A23. \doi{10.1051/0004-6361/200913852}

\bibitem[{Dijkstra et~al(2018)Dijkstra, van~de Loosdrecht, Schomaker, and Wiering}]{dijkstra2018hyperspectral}
Dijkstra K, van~de Loosdrecht J, Schomaker LRB, et~al (2018) Hyperspectral demosaicking and crosstalk correction using deep learning. Machine Vision and Applications 30(1). \doi{10.1007/s00138-018-0965-4}, \urlprefix\url{https://doi.org/10.1007/s00138-018-0965-4}

\bibitem[{{Dotto} et~al(2024){Dotto}, {Deshapriya}, {Gai}, {Hasselmann}, {Mazzotta Epifani}, {Poggiali}, {Rossi}, {Zanotti}, {Zinzi}, {Bertini}, {Brucato}, {Dall'Ora}, {Della Corte}, {Ivanovski}, {Lucchetti}, {Pajola}, {Amoroso}, {Barnouin}, {Campo Bagatin}, {Capannolo}, {Caporali}, {Ceresoli}, {Chabot}, {Cheng}, {Cremonese}, {Fahnestock}, {Farnham}, {Ferrari}, {Gomez Casajus}, {Gramigna}, {Hirabayashi}, {Ieva}, {Impresario}, {Jutzi}, {Lasagni Manghi}, {Lavagna}, {Li}, {Lombardo}, {Modenini}, {Palumbo}, {Perna}, {Pirrotta}, {Raducan}, {Richardson}, {Rivkin}, {Stickle}, {Sunshine}, {Tortora}, {Tusberti}, and {Zannoni}}]{2024Natur.627..505D}
{Dotto} E, {Deshapriya} JDP, {Gai} I, et~al (2024) {The Dimorphos ejecta plume properties revealed by LICIACube}. \nat 627(8004):505--509. \doi{10.1038/s41586-023-06998-2}

\bibitem[{{ESA SPICE Service}(2025)}]{esa_spice_operational_skd_2025}
{ESA SPICE Service} (2025) Operational spice kernel dataset. Data set, SPICE Service, European Space Agency, \doi{10.57780/esa-k25x2cv}, operational SPICE Kernel Dataset (SKD) for the HERA mission

\bibitem[{{Esposito} and {Marchi}(2015)}]{2015MetroAeroSpace..547E}
{Esposito} M, {Marchi} AZ (2015) {HyperCube the Intelligent Hyperspectral Imager}. In: 2015 IEEE Metrology for Aerospace (MetroAeroSpace). IEEE, Benevento, Italy, pp 547--550, \doi{10.1109/MetroAeroSpace.2015.7180716}

\bibitem[{{Esposito} and {Marchi}(2019)}]{2019SPIE11180E..71E}
{Esposito} M, {Marchi} AZ (2019) {In-Orbit Demonstration of the First Hyperspectral Imager for Nanosatellites}. In: {Karafolas} N, {Sodnik} Z, {Cugny} B (eds) International Conference on Space Optics — ICSO 2018, vol 11180. SPIE, Chania, Greece, p~71, \doi{10.1117/12.2535991}

\bibitem[{{Esposito} et~al(2018){Esposito}, {Conticello}, {Vercruyssen}, {Van Dijk}, {Manzillo}, {Delauré}, {Benhadj} et~al}]{2018AIAA....Esposito}
{Esposito} M, {Conticello} SS, {Vercruyssen} N, et~al (2018) {Demonstration in Space of a Smart Hyperspectral Imager for Nanosatellites}. In: 32nd Annual AIAA/USU Conference on Small Satellites

\bibitem[{{Farnham} et~al(2025){Farnham}, {Sunshine}, {Hirabayashi}, {Ernst}, {Daly}, {Agrusa}, {Barnouin}, {Li}, {Kumamoto}, {Syal}, {Wiggins}, {Bjonnes}, {Stickle}, {Raducan}, {Cheng}, {Glenar}, {Lolachi}, {Stubbs}, {Fahnstock}, {Amoroso}, {Bertini}, {Brucato}, {Capannolo}, {Cremonese}, {Dall'Ora}, {Della Corte}, {Deshapriya}, {Dotto}, {Gai}, {Hasselmann}, {Ieva}, {Impresario}, {Ivanovski}, {Lavagna}, {Lucchetti}, {Marzari}, {Epifani}, {Modenini}, {Pajola}, {Palumbo}, {Pirrotta}, {Poggiali}, {Rossi}, {Tortora}, {Zannoni}, {Zanotti}, and {Zinzi}}]{2025PSJ.....6..155F}
{Farnham} TL, {Sunshine} JM, {Hirabayashi} M, et~al (2025) {High-speed Boulders and the Debris Field in DART Ejecta}. \psj 6(7):155. \doi{10.3847/PSJ/addd1a}, {\href{https://arxiv.org/abs/2506.16694}{{arXiv:2506.16694}}} {[astro-ph.EP]}

\bibitem[{{Foglia Manzillo} et~al(2022){Foglia Manzillo}, {Esposito}, {van Dijk}, {Vercruyssen}, {Koeleman}, {Mangini}, {Gatti}, {Harpur}, {Castiglione}, {Goldberg}, and {Carnelli}}]{hshiac}
{Foglia Manzillo} P, {Esposito} M, {van Dijk} CN, et~al (2022) {The spectral imager for the planetary defence mission HERA}. In: 73rd International Astronautical Congress (IAC), Paris, France, paper presented at the 73rd IAC, 18--22 September 2022

\bibitem[{{Grieve} and {Pesonen}(1992)}]{1992Tectp.216....1G}
{Grieve} RAF, {Pesonen} LJ (1992) {The terrestrial impact cratering record.} Tectonophysics 216:1--30. \doi{10.1016/0040-1951(92)90152-V}

\bibitem[{{Ieva} et~al(2022){Ieva}, {Mazzotta Epifani}, {Perna}, {Dall'Ora}, {Petropoulou}, {Deshapriya}, {Hasselmann}, {Rossi}, {Poggiali}, {Brucato}, {Pajola}, {Lucchetti}, {Ivanovski}, {Palumbo}, {Della Corte}, {Zinzi}, {Rivkin}, {Thomas}, {de Le{\'o}n}, {Dotto}, {Amoroso}, {Bertini}, {Capannolo}, {Cotugno}, {Cremonese}, {Di Tana}, {Gai}, {Impresario}, {Lavagna}, {Meneghin}, {Miglioretti}, {Modenini}, {Pirrotta}, {Simioni}, {Simonetti}, {Tortora}, {Zannoni}, and {Zanotti}}]{2022PSJ.....3..183I}
{Ieva} S, {Mazzotta Epifani} E, {Perna} D, et~al (2022) {Spectral Rotational Characterization of the Didymos System prior to the DART Impact}. \psj 3(8):183. \doi{10.3847/PSJ/ac7f34}

\bibitem[{{Ishiguro} et~al(2007){Ishiguro}, {Hiroi}, {Tholen}, {Sasaki}, {Ueda}, {Nimura}, {Abe}, {Clark}, {Yamamoto}, {Yoshida}, {Nakamura}, {Hirata}, {Miyamoto}, {Yokota}, {Hashimoto}, {Kubota}, {Nakamura}, {Gaskell}, and {Saito}}]{2007M&PS...42.1791I}
{Ishiguro} M, {Hiroi} T, {Tholen} DJ, et~al (2007) {Global mapping of the degree of space weathering on asteroid 25143 Itokawa by Hayabusa/AMICA observations}. \maps 42(10):1791--1800. \doi{10.1111/j.1945-5100.2007.tb00538.x}

\bibitem[{{Izidoro} et~al(2013){Izidoro}, {de Souza Torres}, {Winter}, and {Haghighipour}}]{2013ApJ...767...54I}
{Izidoro} A, {de Souza Torres} K, {Winter} OC, et~al (2013) {A Compound Model for the Origin of Earth's Water}. \apj 767(1):54. \doi{10.1088/0004-637X/767/1/54}, {\href{https://arxiv.org/abs/1302.1233}{{arXiv:1302.1233}}} {[astro-ph.EP]}

\bibitem[{{Koga} et~al(2018){Koga}, {Sugita}, {Kamata}, {Ishiguro}, {Hiroi}, {Tatsumi}, and {Sasaki}}]{2018Icar..299..386K}
{Koga} SC, {Sugita} S, {Kamata} S, et~al (2018) {Spectral decomposition of asteroid Itokawa based on principal component analysis}. \icarus 299:386--395. \doi{10.1016/j.icarus.2017.08.016}

\bibitem[{{Korda} and {Kohout}(2024)}]{2024PSJ.....5...85K}
{Korda} D, {Kohout} T (2024) {Silicate Mineralogy from Vis{\textendash}NIR Reflectance Spectra}. \psj 5(4):85. \doi{10.3847/PSJ/ad2685}

\bibitem[{{Korda} et~al(2023){Korda}, {Penttil{\"a}}, {Klami}, and {Kohout}}]{2023A&A...669A.101K}
{Korda} D, {Penttil{\"a}} A, {Klami} A, et~al (2023) {Neural network for determining an asteroid mineral composition from reflectance spectra}. \aap 669:A101. \doi{10.1051/0004-6361/202243886}, {\href{https://arxiv.org/abs/2210.01006}{{arXiv:2210.01006}}} {[astro-ph.EP]}

\bibitem[{{Kuramoto} et~al(2022){Kuramoto}, {Kawakatsu}, {Fujimoto}, {Araya}, {Barucci}, {Genda}, {Hirata}, {Ikeda}, {Imamura}, {Helbert}, {Kameda}, {Kobayashi}, {Kusano}, {Lawrence}, {Matsumoto}, {Michel}, {Miyamoto}, {Morota}, {Nakagawa}, {Nakamura}, {Ogawa}, {Otake}, {Ozaki}, {Russell}, {Sasaki}, {Sawada}, {Senshu}, {Tachibana}, {Terada}, {Ulamec}, {Usui}, {Wada}, {Watanabe}, and {Yokota}}]{2022EP&S...74...12K}
{Kuramoto} K, {Kawakatsu} Y, {Fujimoto} M, et~al (2022) {Martian moons exploration MMX: sample return mission to Phobos elucidating formation processes of habitable planets}. Earth, Planets and Space 74(1):12. \doi{10.1186/s40623-021-01545-7}

\bibitem[{{Lazzarin} et~al(2023){Lazzarin}, {La Forgia}, {Migliorini}, {Farina}, {Frattin}, and {Ochner}}]{2023LPICo2806.2080L}
{Lazzarin} M, {La Forgia} F, {Migliorini} A, et~al (2023) {Rotationally-Resolved Characterization of the Near-Earth Didymos-Dimorphos Binary System After the NASA/DART Impact}. In: 54th Lunar and Planetary Science Conference, p 2080

\bibitem[{{MacLennan} et~al(2024){MacLennan}, {Emery}, {Lucas}, {MCClure}, and {Lindsay}}]{2024M&PS...59.1329M}
{MacLennan} EM, {Emery} JP, {Lucas} MP, et~al (2024) {Space weathering, grain size, and metamorphic heating effects on ordinary chondrite spectral reflectance parameters}. \maps 59(6):1329--1352. \doi{10.1111/maps.14150}

\bibitem[{{Mahlke} et~al(2022){Mahlke}, {Carry}, and {Mattei}}]{2022A&A...665A..26M}
{Mahlke} M, {Carry} B, {Mattei} PA (2022) {Asteroid taxonomy from cluster analysis of spectrometry and albedo}. \aap 665:A26. \doi{10.1051/0004-6361/202243587}, {\href{https://arxiv.org/abs/2203.11229}{{arXiv:2203.11229}}} {[astro-ph.EP]}

\bibitem[{{Marchi} et~al(2012){Marchi}, {Paolicchi}, and {Richardson}}]{2012MNRAS.421....2M}
{Marchi} S, {Paolicchi} P, {Richardson} DC (2012) {Collisional evolution and reddening of asteroid surfaces - I. The problem of conflicting time-scales and the role of size-dependent effects}. \mnras 421(1):2--8. \doi{10.1111/j.1365-2966.2011.19952.x}, {\href{https://arxiv.org/abs/1110.2503}{{arXiv:1110.2503}}} {[astro-ph.EP]}

\bibitem[{{Marty} et~al(2016){Marty}, {Avice}, {Sano}, {Altwegg}, {Balsiger}, {H{\"a}ssig}, {Morbidelli}, {Mousis}, and {Rubin}}]{2016E&PSL.441...91M}
{Marty} B, {Avice} G, {Sano} Y, et~al (2016) {Origins of volatile elements (H, C, N, noble gases) on Earth and Mars in light of recent results from the ROSETTA cometary mission}. Earth and Planetary Science Letters 441:91--102. \doi{10.1016/j.epsl.2016.02.031}

\bibitem[{{McLean}(2008)}]{2008eiad.book.....M}
{McLean} IS (2008) {Electronic Imaging in Astronomy: Detectors and Instrumentation (Second Edition)}

\bibitem[{{Michel} et~al(2016){Michel}, {Cheng}, {K{\"u}ppers}, {Pravec}, {Blum}, {Delbo}, {Green}, {Rosenblatt}, {Tsiganis}, {Vincent}, {Biele}, {Ciarletti}, {H{\'e}rique}, {Ulamec}, {Carnelli}, {Galvez}, {Benner}, {Naidu}, {Barnouin}, {Richardson}, {Rivkin}, {Scheirich}, {Moskovitz}, {Thirouin}, {Schwartz}, {Campo Bagatin}, and {Yu}}]{2016AdSpR..57.2529M}
{Michel} P, {Cheng} A, {K{\"u}ppers} M, et~al (2016) {Science case for the Asteroid Impact Mission (AIM): A component of the Asteroid Impact \& Deflection Assessment (AIDA) mission}. Advances in Space Research 57(12):2529--2547. \doi{10.1016/j.asr.2016.03.031}

\bibitem[{{Michel} et~al(2022){Michel}, {K{\"u}ppers}, {Bagatin}, {Carry}, {Charnoz}, {de Leon}, {Fitzsimmons}, {Gordo}, {Green}, {H{\'e}rique}, {Juzi}, {Karatekin}, {Kohout}, {Lazzarin}, {Murdoch}, {Okada}, {Palomba}, {Pravec}, {Snodgrass}, {Tortora}, {Tsiganis}, {Ulamec}, {Vincent}, {W{\"u}nnemann}, {Zhang}, {Raducan}, {Dotto}, {Chabot}, {Cheng}, {Rivkin}, {Barnouin}, {Ernst}, {Stickle}, {Richardson}, {Thomas}, {Arakawa}, {Miyamoto}, {Nakamura}, {Sugita}, {Yoshikawa}, {Abell}, {Asphaug}, {Ballouz}, {Bottke}, {Lauretta}, {Walsh}, {Martino}, and {Carnelli}}]{michel2022psj}
{Michel} P, {K{\"u}ppers} M, {Bagatin} AC, et~al (2022) {The ESA Hera Mission: Detailed Characterization of the DART Impact Outcome and of the Binary Asteroid (65803) Didymos}. \psj 3(7):160. \doi{10.3847/PSJ/ac6f52}

\bibitem[{Mihoubi(2018)}]{mihoubi2018snapshot}
Mihoubi S (2018) Snapshot multispectral image demosaicing and classification. PhD thesis, Université de Lille, \urlprefix\url{https://hal.archives-ouvertes.fr/tel-01953493}, english

\bibitem[{{Morate} et~al(2023){Morate}, {Popescu}, {Licandro}, {Tinaut-Ruano}, {Tatsumi}, and {de Le{\'o}n}}]{2023MNRAS.519.1677M}
{Morate} D, {Popescu} M, {Licandro} J, et~al (2023) {Mineralogical analysis of 14 PHAs from ViNOS data}. \mnras 519(2):1677--1687. \doi{10.1093/mnras/stac3530}

\bibitem[{Morbidelli and Vokrouhlick\'y(2003)}]{2003Icar..163..120M}
Morbidelli A, Vokrouhlick\'y D (2003) The yarkovsky-driven origin of near-earth asteroids. Icarus 163:120--134. \doi{10.1016/S0019-1035(03)00047-2}

\bibitem[{Morbidelli et~al(2002)Morbidelli, {Bottke, W.F.Jr.}, Froeschl\'e, and Michel}]{2002aste.book..409M}
Morbidelli A, {Bottke, W.F.Jr.}, Froeschl\'e C, et~al (2002) Origin and {E}volution of {N}ear-{E}arth {O}bjects. In: {Bottke, W. F. Jr.}, Cellino A, Paolichi P, et~al (eds) Asteroids III. University of Arizona Press, Tucson, p 409--422

\bibitem[{{Moskovitz} et~al(2024){Moskovitz}, {Thomas}, {Pravec}, {Lister}, {Polakis}, {Osip}, {Kareta}, {Ro{\.z}ek}, {Chesley}, {Naidu}, {Scheirich}, {Ryan}, {Ryan}, {Skiff}, {Snodgrass}, {Knight}, {Rivkin}, {Chabot}, {Ayvazian}, {Belskaya}, {Benkhaldoun}, {Berte{\c{s}}teanu}, {Bonavita}, {Bressi}, {Brucker}, {Burgdorf}, {Burkhonov}, {Burt}, {Contreras}, {Chatelain}, {Choi}, {Daily}, {de Le{\'o}n}, {Ergashev}, {Farnham}, {Fatka}, {Ferrais}, {Geier}, {Gomez}, {Greenstreet}, {Gr{\"o}ller}, {Hergenrother}, {Holt}, {Hornoch}, {Hus{\'a}rik}, {Inasaridze}, {Jehin}, {Khalouei}, {Eluo}, {Kim}, {Krugly}, {Ku{\v{c}}{\'a}kov{\'a}}, {Ku{\v{s}}nir{\'a}k}, {Larsen}, {Lee}, {Lejoly}, {Licandro}, {Longa-Pe{\~n}a}, {Mastaler}, {McCully}, {Moon}, {Morrell}, {Nath}, {Oszkiewicz}, {Parrott}, {Phillips}, {Popescu}, {Pray}, {Prodan}, {Rabus}, {Read}, {Reva}, {Roark}, {Santana-Ros}, {Scotti}, {Tatara}, {Thirouin}, {Tholen}, {Troianskyi}, {Tubbiolo}, and {Villa}}]{2024PSJ.....5...35M}
{Moskovitz} N, {Thomas} C, {Pravec} P, et~al (2024) {Photometry of the Didymos System across the DART Impact Apparition}. \psj 5(2):35. \doi{10.3847/PSJ/ad0e74}, {\href{https://arxiv.org/abs/2311.01971}{{arXiv:2311.01971}}} {[astro-ph.EP]}

\bibitem[{{Naidu} et~al(2024){Naidu}, {Chesley}, {Moskovitz}, {Thomas}, {Meyer}, {Pravec}, {Scheirich}, {Farnocchia}, {Scheeres}, {Brozovic}, {Benner}, {Rivkin}, and {Chabot}}]{2024PSJ.....5...74N}
{Naidu} SP, {Chesley} SR, {Moskovitz} N, et~al (2024) {Orbital and Physical Characterization of Asteroid Dimorphos Following the DART Impact}. \psj 5(3):74. \doi{10.3847/PSJ/ad26e7}

\bibitem[{{Pieters} and {Noble}(2016)}]{2016JGRE..121.1865P}
{Pieters} CM, {Noble} SK (2016) {Space weathering on airless bodies}. Journal of Geophysical Research (Planets) 121(10):1865--1884. \doi{10.1002/2016JE005128}

\bibitem[{{Polishook} et~al(2023){Polishook}, {DeMeo}, {Burt}, {Thomas}, {Rivkin}, {Sanchez}, and {Reddy}}]{2023PSJ.....4..229P}
{Polishook} D, {DeMeo} FE, {Burt} BJ, et~al (2023) {Near-IR Spectral Observations of the Didymos System: Daily Evolution Before and After the DART Impact Indicates that Dimorphos Originated from Didymos}. \psj 4(12):229. \doi{10.3847/PSJ/ad08ae}, {\href{https://arxiv.org/abs/2311.00421}{{arXiv:2311.00421}}} {[astro-ph.EP]}

\bibitem[{{Popescu} et~al(2019){Popescu}, {Vaduvescu}, {de Le{\'o}n}, {Gherase}, {Licandro}, {Boac{\u{a}}}, {{\c{S}}onka}, {Ashley}, {Mo{\v{c}}nik}, {Morate}, {Predatu}, {De Pr{\'a}}, {Fari{\~n}a}, {Stoev}, {D{\'\i}az Alfaro}, {Ordonez-Etxeberria}, {L{\'o}pez-Mart{\'\i}nez}, and {Errmann}}]{2019A&A...627A.124P}
{Popescu} M, {Vaduvescu} O, {de Le{\'o}n} J, et~al (2019) {Near-Earth asteroids spectroscopic survey at Isaac Newton Telescope}. \aap 627:A124. \doi{10.1051/0004-6361/201935006}, {\href{https://arxiv.org/abs/1905.12997}{{arXiv:1905.12997}}} {[astro-ph.EP]}

\bibitem[{{Pravec} and {Harris}(2000)}]{2000Icar..148...12P}
{Pravec} P, {Harris} AW (2000) {Fast and Slow Rotation of Asteroids}. \icarus 148(1):12--20. \doi{10.1006/icar.2000.6482}

\bibitem[{{Pravec} et~al(2022){Pravec}, {Thomas}, {Rivkin}, {Scheirich}, {Moskovitz}, {Knight}, {Snodgrass}, {de Le{\'o}n}, {Licandro}, {Popescu}, {Thirouin}, {F{\"o}hring}, {Chandler}, {Oldroyd}, {Trujillo}, {Howell}, {Green}, {Thomas-Osip}, {Sheppard}, {Farnham}, {Mazzotta Epifani}, {Dotto}, {Ieva}, {Dall'Ora}, {Kokotanekova}, {Carry}, and {Souami}}]{2022PSJ.....3..175P}
{Pravec} P, {Thomas} CA, {Rivkin} AS, et~al (2022) {Photometric Observations of the Binary Near-Earth Asteroid (65803) Didymos in 2015-2021 Prior to DART Impact}. \psj 3(7):175. \doi{10.3847/PSJ/ac7be1}

\bibitem[{{Pravec} et~al(2024){Pravec}, {Meyer}, {Scheirich}, {Scheeres}, {Benson}, and {Agrusa}}]{2024Icar..41816138P}
{Pravec} P, {Meyer} AJ, {Scheirich} P, et~al (2024) {Rotational lightcurves of Dimorphos and constraints on its post-DART impact spin state}. \icarus 418:116138. \doi{10.1016/j.icarus.2024.116138}

\bibitem[{{Raducan} et~al(2024){Raducan}, {Jutzi}, {Cheng}, {Zhang}, {Barnouin}, {Collins}, {Daly}, {Davison}, {Ernst}, {Farnham}, {Ferrari}, {Hirabayashi}, {Kumamoto}, {Michel}, {Murdoch}, {Nakano}, {Pajola}, {Rossi}, {Agrusa}, {Barbee}, {Syal}, {Chabot}, {Dotto}, {Fahnestock}, {Hasselmann}, {Herreros}, {Ivanovski}, {Li}, {Lucchetti}, {Luther}, {Orm{\"o}}, {Owen}, {Pravec}, {Rivkin}, {Robin}, {S{\'a}nchez}, {Tusberti}, {W{\"u}nnemann}, {Zinzi}, {Epifani}, {Manzoni}, and {May}}]{2024NatAs...8..445R}
{Raducan} SD, {Jutzi} M, {Cheng} AF, et~al (2024) {Physical properties of asteroid Dimorphos as derived from the DART impact}. Nature Astronomy 8:445--455. \doi{10.1038/s41550-024-02200-3}

\bibitem[{{Rivkin} et~al(2021){Rivkin}, {Chabot}, {Stickle}, {Thomas}, {Richardson}, {Barnouin}, {Fahnestock}, {Ernst}, {Cheng}, {Chesley}, {Naidu}, {Statler}, {Barbee}, {Agrusa}, {Moskovitz}, {Terik Daly}, {Pravec}, {Scheirich}, {Dotto}, {Della Corte}, {Michel}, {K{\"u}ppers}, {Atchison}, and {Hirabayashi}}]{2021PSJ.....2..173R}
{Rivkin} AS, {Chabot} NL, {Stickle} AM, et~al (2021) {The Double Asteroid Redirection Test (DART): Planetary Defense Investigations and Requirements}. \psj 2(5):173. \doi{10.3847/PSJ/ac063e}

\bibitem[{{Rivkin} et~al(2023){Rivkin}, {Thomas}, {Wong}, {Rozitis}, {de Le{\'o}n}, {Holler}, {Milam}, {Howell}, {Hammel}, {Arredondo}, {Brucato}, {Epifani}, {Ieva}, {La Forgia}, {Lucas}, {Lucchetti}, {Pajola}, {Poggiali}, {Sunshine}, and {Trigo-Rodr{\'\i}guez}}]{2023PSJ.....4..214R}
{Rivkin} AS, {Thomas} CA, {Wong} I, et~al (2023) {Near to Mid-infrared Spectroscopy of (65803) Didymos as Observed by JWST: Characterization Observations Supporting the Double Asteroid Redirection Test}. \psj 4(11):214. \doi{10.3847/PSJ/ad04d8}, {\href{https://arxiv.org/abs/2310.11168}{{arXiv:2310.11168}}} {[astro-ph.EP]}

\bibitem[{{Scheirich} et~al(2024){Scheirich}, {Pravec}, {Meyer}, {Agrusa}, {Richardson}, {Chesley}, {Naidu}, {Thomas}, and {Moskovitz}}]{2024PSJ.....5...17S}
{Scheirich} P, {Pravec} P, {Meyer} AJ, et~al (2024) {Dimorphos Orbit Determination from Mutual Events Photometry}. \psj 5(1):17. \doi{10.3847/PSJ/ad12cf}, {\href{https://arxiv.org/abs/2403.02804}{{arXiv:2403.02804}}} {[astro-ph.EP]}

\bibitem[{{Statler} et~al(2022){Statler}, {Raducan}, {Barnouin}, {DeCoster}, {Chesley}, {Barbee}, {Agrusa}, {Cambioni}, {Cheng}, {Dotto}, {Eggl}, {Fahnestock}, {Ferrari}, {Graninger}, {Herique}, {Herreros}, {Hirabayashi}, {Ivanovski}, {Jutzi}, {Karatekin}, {Lucchetti}, {Luther}, {Makadia}, {Marzari}, {Michel}, {Murdoch}, {Nakano}, {Orm{\"o}}, {Pajola}, {Rivkin}, {Rossi}, {S{\'a}nchez}, {Schwartz}, {Soldini}, {Souami}, {Stickle}, {Tortora}, {Trigo-Rodr{\'\i}guez}, {Venditti}, {Vincent}, and {W{\"u}nnemann}}]{2022PSJ.....3..244S}
{Statler} TS, {Raducan} SD, {Barnouin} OS, et~al (2022) {After DART: Using the First Full-scale Test of a Kinetic Impactor to Inform a Future Planetary Defense Mission}. \psj 3(10):244. \doi{10.3847/PSJ/ac94c1}, {\href{https://arxiv.org/abs/2209.11873}{{arXiv:2209.11873}}} {[astro-ph.EP]}

\bibitem[{{Suzuki} et~al(2018){Suzuki}, {Yamada}, {Kouyama}, {Tatsumi}, {Kameda}, {Honda}, {Sawada}, {Ogawa}, {Morota}, {Honda}, {Sakatani}, {Hayakawa}, {Yokota}, {Yamamoto}, and {Sugita}}]{2018Icar..300..341S}
{Suzuki} H, {Yamada} M, {Kouyama} T, et~al (2018) {Initial inflight calibration for Hayabusa2 optical navigation camera (ONC) for science observations of asteroid Ryugu}. \icarus 300:341--359. \doi{10.1016/j.icarus.2017.09.011}

\bibitem[{{Tatsumi} et~al(2021){Tatsumi}, {Popescu}, {Campins}, {de Le{\'o}n}, {Garc{\'\i}a}, {Licandro}, {Simon}, {Kaplan}, {DellaGiustina}, {Golish}, and {Lauretta}}]{2021MNRAS.508.2053T}
{Tatsumi} E, {Popescu} M, {Campins} H, et~al (2021) {Widely distributed exogenic materials of varying compositions and morphologies on asteroid (101955) Bennu}. \mnras 508(2):2053--2070. \doi{10.1093/mnras/stab2548}, {\href{https://arxiv.org/abs/2109.01449}{{arXiv:2109.01449}}} {[astro-ph.EP]}

\bibitem[{{Thomas} et~al(2023){Thomas}, {Naidu}, {Scheirich}, {Moskovitz}, {Pravec}, {Chesley}, {Rivkin}, {Osip}, {Lister}, {Benner}, {Brozovi{\'c}}, {Contreras}, {Morrell}, {Ro{\.Z}ek}, {Ku{\v{s}}nir{\'a}k}, {Hornoch}, {Mages}, {Taylor}, {Seymour}, {Snodgrass}, {J{\o}rgensen}, {Dominik}, {Skiff}, {Polakis}, {Knight}, {Farnham}, {Giorgini}, {Rush}, {Bellerose}, {Salas}, {Armentrout}, {Watts}, {Busch}, {Chatelain}, {Gomez}, {Greenstreet}, {Phillips}, {Bonavita}, {Burgdorf}, {Khalouei}, {Longa-Pe{\~n}a}, {Rabus}, {Sajadian}, {Chabot}, {Cheng}, {Ryan}, {Ryan}, {Holt}, and {Agrusa}}]{2023Natur.616..448T}
{Thomas} CA, {Naidu} SP, {Scheirich} P, et~al (2023) {Orbital period change of Dimorphos due to the DART kinetic impact}. \nat 616(7957):448--451. \doi{10.1038/s41586-023-05805-2}, {\href{https://arxiv.org/abs/2303.02077}{{arXiv:2303.02077}}} {[astro-ph.EP]}

\bibitem[{Tsagkatakis et~al(2019)Tsagkatakis, Bloemen, Geelen, Jayapala, and Tsakalides}]{Tsagkatakis}
Tsagkatakis G, Bloemen M, Geelen B, et~al (2019) Graph and rank regularized matrix recovery for snapshot spectral image demosaicing. IEEE Transactions on Computational Imaging 5(2):301--316. \doi{10.1109/TCI.2018.2888989}

\bibitem[{{Vincent} et~al(2012){Vincent}, {Besse}, {Marchi}, {Sierks}, {Massironi}, {OSIRIS Team}, {A'Hearn}, {Angrilli}, {Barbieri}, {Barucci}, {Bertaux}, {Cremonese}, {Da Deppo}, {Davidsson}, {Debei}, {De Cecco}, {Fornasier}, {Fulle}, {Groussin}, {Gutierrez}, {Hviid}, {Ip}, {Jorda}, {Keller}, {Koschny}, {Knollenberg}, {Kramm}, {Kuehrt}, {Lamy}, {Lara}, {Lazzarin}, {Lopez-Moreno}, {Marzari}, {Michalik}, {Naletto}, {Rickman}, {Rodrigo}, {Sabau}, {Thomas}, and {Wenzel}}]{2012P&SS...66...79V}
{Vincent} JB, {Besse} S, {Marchi} S, et~al (2012) {Physical properties of craters on asteroid (21) Lutetia}. Planetary and Space Science 66(1):79--86. \doi{10.1016/j.pss.2011.12.025}

\bibitem[{Virtanen et~al(2020)Virtanen, Gommers, Oliphant, Haberland, Reddy, Cournapeau, Burovski, Peterson, Weckesser, Bright, {van der Walt}, Brett, Wilson, Millman, Mayorov, Nelson, Jones, Kern, Larson, Carey, Polat, Feng, Moore, {VanderPlas}, Laxalde, Perktold, Cimrman, Henriksen, Quintero, Harris, Archibald, Ribeiro, Pedregosa, {van Mulbregt}, and {SciPy 1.0 Contributors}}]{2020SciPy-NMeth}
Virtanen P, Gommers R, Oliphant TE, et~al (2020) {{SciPy} 1.0: Fundamental Algorithms for Scientific Computing in Python}. Nature Methods 17:261--272. \doi{10.1038/s41592-019-0686-2}

\bibitem[{{Walsh} et~al(2008){Walsh}, {Richardson}, and {Michel}}]{2008Natur.454..188W}
{Walsh} KJ, {Richardson} DC, {Michel} P (2008) {Rotational breakup as the origin of small binary asteroids}. \nat 454(7201):188--191. \doi{10.1038/nature07078}

\end{thebibliography}

\end{document}